\documentclass[aps,floats,superscriptaddress]{revtex4}
\usepackage{amsmath}
\usepackage{graphicx}

\def\moy#1{\left\langle #1 \right\rangle}
\def\ppint{P\int}
\def\Im{\hbox{Im}}
\def\Tr{\text{Tr}}
\def\sgn{\text{sgn}\,}

\def\fig#1#2{\includegraphics[height=#1]{#2}}

\begin{document}

\title{Quantum Fluctuations of a Nearly Critical Heisenberg Spin Glass}

\author{A. Georges}
\affiliation{Laboratoire de Physique Th{\'e}orique, Ecole Normale
Sup{\'e}rieure, 24 Rue Lhomond 75005 Paris FRANCE}
\author{O. Parcollet}
\affiliation{Center for Material Theory,
Department of Physics and Astronomy, Rutgers University, Piscataway, NJ
08854 USA}
\author{S. Sachdev}
\affiliation{Department of Physics, Yale University, New Haven, CT
06520 USA}
\begin{abstract}
We describe the interplay of quantum and thermal fluctuations in
the infinite-range Heisenberg spin glass. This model is
generalized to $SU(N)$ symmetry, and we describe the phase
diagram as a function of the spin $S$ and the temperature $T$.
The model is solved in the large $N$ limit and certain universal
critical properties are shown to hold to all orders in $1/N$. For
large $S$, the ground state is a spin glass, but quantum effects
are crucial in determining the low $T$ thermodynamics: we find a
specific heat linear in $T$ and a local spectral density of spin
excitations, $\chi^{\prime\prime}_{loc} ( \omega ) \sim \omega$
for a spin glass state which is marginally stable to fluctuations
in the replicon modes. For small $S$, the spin-glass order is
fragile, and a spin-liquid state with $\chi^{\prime\prime}_{loc}
\sim \tanh (\omega/2T)$ dominates the properties over a
significant range of $T$ and $\omega$. We argue that the latter
state may be relevant in understanding the properties of
strongly-disordered transition metal and rare earth compounds.
\end{abstract}

\maketitle

The study of intermetallic compounds of the transition metals and
rare earths has been a subject at the forefront of condensed
matter physics for some time now \cite{itp,coleman}. 
A rich and complex variety of
behaviors is observed in low temperature electrical and magnetic
measurements, much of which lacks a comprehensive theoretical
description. The complexity arises from the dominant role played
by the local magnetic moments on the $d$ and $f$ orbitals and
their interactions with each other and the itinerant charge
carriers.

It is convenient to begin our discussion in a phase with
well-established magnetic order, in which
each magnetic
moment is effectively static. This static moment could be
polarized in a regular manner (as in a commensurate
antiferromagnet or an incommensurate spin density wave), or point
in random directions (as in a spin glass state). In most realistic
systems, the magnetic moment is either quite small, or has
averaged to zero by dynamic quantum fluctuations: so it is
useful to consider mechanisms which reduce the magnetic moment,
and eventually cause it to vanish at a quantum phase transition to
some paramagnetic state. Two distinct routes to such a quantum
phase transition can be envisaged, and, we believe, the interplay
between them is at the heart of the complexity of the problem. In
the first route, originally discussed by Doniach \cite{Doniach}, the moment is
quenched by Kondo screening by the itinerant electrons: theories
of such quantum critical points have been proposed\cite{Hertz,Millis,SRO,AG}
in which the
predominant role of the itinerancy is to overdamp the collective
magnetic excitations. In the second route, the exchange
interactions between the moments play a more fundamental role:
a pair of spins interacting with an antiferromagnetic exchange
prefers to form a singlet valence bond, and the proliferation of
such singlets can destroy the magnetic order. Analytic theories
for such transitions has been made mainly for systems without
quenched disorder\cite{SachdevBook}.
Simple models of crossovers between these two
routes have also been presented\cite{SCS,GS,AMS}.

This paper will present a detailed study of the second route
to destruction of magnetic order for the case of a strongly random
system with spin glass magnetic order. There are a number of
motivations for
focusing on random systems. First, randomness is
inevitably present in all materials, and it is clear that it
strongly perturbs the low temperature properties. Spin glass order
is present in a number of systems, while others appear to be in the
vicinity of such a state. Finally, a technical motivation is in
the structure of the mean-field theory we shall present: it builds
in important feedback effect between the inter-site magnetic
correlations and the single-site spin dynamics, and this is
crucial to all the non-trivial spin correlations we shall
describe. Such a feedback is absent in previous studies of the
magnetic quantum critical point, and it has been argued that this
is an important limitation for them\cite{AG,AMS}.
A different route to
incorporating these feedback effects has been taken in some recent
studies\cite{Si}; however, they discuss only the paramagnetic state of
their model, and the extent to which magnetically ordered states
preempt their results remains to be clarified.

This paper is organized as follows : In Section \ref{outline}, we
present our spin glass model and give an outline of our results,
including the phase diagram.  Section
\ref{ParamagneticPhases} is devoted to the nature of the
paramagnetic solutions, and more specifically to the quantum
critical regime. Section \ref{SpinGlassPhases} is devoted to the
spin glass phase and to the various regimes within this phase, as
a function of temperature and of the size of the spin. The
Appendices contain technical details and some additional results
on the quantum rotor and Ising spin glasses of
Ref.~\onlinecite{rsy}.

\section{Model and outline of the results}\label{outline}

The numerous recent studies of quantum
fluctuations in spin glasses \cite{RevueBhatt},
have focused either on infinite-range models of Ising
and rotor models \cite{millerhuse,rsy}
or models in low dimensions which flow to
strong disorder fixed points \cite{dsf,Pich,Motrunich}.
Here we shall continue the study of infinite range models,
but will consider a model of Heisenberg spins :  in this case,
the path integral for each spin has a important Berry phase term
which imposes the spin
commutation relations.
As we will see, this leads to a great deal of
new physics\cite{SachdevYe} and non-trivial dynamic spin correlations
even in the spin glass state.
More specifically, we present a complete solution of
the quantum Heisenberg spin glass on a fully
connected lattice of $\cal N$ sites with strong Gaussian disorder,
both in the paramagnetic and the glassy phase,
when the spin symmetry group is extended from $SU(2)$ to $SU(N)$ and the
large-$N$ limit is taken.
In the limit of large connectivity, (dynamical) mean-field techniques
apply and the
model can be reduced to the study of a self-consistent single-site
problem, which  is however
still highly non-trivial  because of quantum effects.
The large-$N$ limit is instrumental in allowing for an explicit
solution. Nevertheless some of our results regarding the quantum
critical regime have been extended beyond the large-N limit.
In a recent publication \cite{NotreLettre}, we summarized the main
results of the present study. Here, we provide detailed derivations
and new results, such as  a full discussion of the paramagnetic
phases and a discussion beyond large-$N$.

The  model considered in this paper is defined by the Hamiltonian:
\begin{equation}\label{DefSachdevYe}
H = \frac{1}{\sqrt{{\cal N}N}} \sum_{i<j} J_{ij}
\vec{S}_{i}\cdot \vec{S}_{j},
\end{equation}
where the magnetic exchange couplings $J_{ij}$ are independent,
quenched random variables  distributed according to a Gaussian distribution
\begin{equation}\label{DefLoiJ}
P (J_{ij}) = \frac{1}{J\sqrt{2\pi}} e^{-J_{ij}^{2}/ (2J^{2})}
\end{equation}
As already pointed out by Bray and Moore \cite{BrayMoore},
after using the replica trick to
average over the disorder \cite{MezardParisiBook}, the mean-field (infinite
dimensional) limit maps the model onto a {\sl self-consistent  single
site model} with the action (in imaginary time $\tau$, with $\beta$ the
inverse temperature)  :
\begin{equation}\label{ActionEffective}
S_{eff} = S_{B} - \frac{J^{2}}{2N} \int_{0}^{\beta} d\tau  d\tau'\,
Q^{ab} (\tau -\tau')
\overrightarrow{S}^{a} (\tau )\cdot \overrightarrow{S}^{b} (\tau')
\end{equation}
and the self-consistency condition
\begin{equation}\label{CondCoherence}
Q^{ab} (\tau -\tau') = \frac{1}{N^{2}}
\moy{
\overrightarrow{S}^{a} (\tau )\cdot  \overrightarrow{S}^{b}(\tau')
}_{S_{eff}}
\end{equation}
where $a,b=1,\cdots,n$ denote the replica indices (the limit
$n\rightarrow 0$ has to be taken later) and $S_{B}$ is the Berry
phase of the spin \cite{SachdevYe}. Due to their time-dependence,
the solution of these mean-field equations remains a very
difficult problem for $N=2$, even in the paramagnetic phase.
Thus, in Ref.~\onlinecite{BrayMoore}, as well as in most
subsequent work \cite{kopec}, the {\it static approximation} was
used, neglecting the $\tau-$dependence of $Q^{ab}(\tau)$. This
approximation may be reasonable in some regimes but prevents a
study of the quantum equilibrium dynamics, and is particularly
inappropriate in the quantum-critical regime. However, this
imaginary time dynamics has been explicitly studied in a Quantum
Monte Carlo simulation {\sl in the paramagnetic phase} with spin
$S=1/2$ by Grempel and Rozenberg \cite{Grempel1}. Recently, we
introduced  a large-$N$ solution of the mean field problem
\cite{NotreLettre}, in which the problem is exactly solvable and,
as explained below, the solution provides a good description of
the physics of the $N=2$ mean field model, to the extent the
latter is understood. More specifically, in the following we will
consider two different types of spin representations for the $SU
(N)$ spins :
\begin{itemize}
\item [a)] {\sl Bosonic} representations : the spin operator $S$ is
represented using Schwinger bosons $b$ by $S_{\alpha \beta}=
b^{\dagger}_{\alpha }b_{\beta } - S\delta_{\alpha \beta }$,
  with the constraint $\sum_{\alpha}
b^{\dagger}_{\alpha }b_{\alpha }= SN$  ($0\leq S$). In
the language of Young tableaux, these representations are described by
 one line of length $SN$.  They
are a natural generalisation of an $SU (2)$ spin of
size $S$.
\item [b)] {\sl Fermionic} representations : the spin operator $S$ is
represented using Abrikosov fermions $f$ by $S_{\alpha \beta}=
f^{\dagger}_{\alpha }f_{\beta } - q_{0} \delta_{\alpha \beta }$,
 with the constraint $\sum_{\alpha}
f^{\dagger}_{\alpha }f_{\alpha }= q_{0}N$  ($0\leq q_{0}\leq 1$). In
the language of Young tableaux, these representations are described by
one column of length $q_{0}N$.
Note that for $SU (2)$, only $S=0$ and $S=1/2$ can be represented in this
manner.
\end{itemize}
In the following, we refer to the model with bosonic (resp.
fermionic) representations
as the bosonic (resp. fermionic) model. In the
fermionic model,  quantum fluctuations are so strong in large-$N$
that the spin glass ordering is destroyed \cite{SachdevYe},
contrary to the bosonic model, where a spin glass phase exists,
as explained below. The two models have different theoretical
interest : if one wants to concentrate on the quantum critical
regime, above the spin glass ordering temperature, one can use the
fermionic model (as e.g. in Ref.~\onlinecite{Slush}). However,
since we are interested in the spin glass phase itself, we will
now focus on the bosonic model. Nevertheless, our results on
the paramagnet will be valid for both cases with only slight
modifications explicitly quoted below.

In the $N\rightarrow \infty$ limit, the mean field self-consistent
model (\ref{ActionEffective}) reduces to
an integral equation for the Green's function of the boson
 $G_{b}^{ab} (\tau) \equiv  - \overline{\moy{T b^{a} (\tau)
b^{\dagger b} (0)}}$ where the bar denotes the average over disorder
and the brackets the thermal average  \cite{SachdevYe} :
\begin{subequations}\label{EqBase}
\begin{align}
(G_{b}^{-1})^{ab} (i\nu_{n}) &=i\nu_{n}\delta_{ab} +
\lambda^{a}\delta_{ab} - \Sigma_{b}^{ab} (i\nu
_{n}) \\
\Sigma^{ab}_{b}(\tau ) &=\ J^{2} \bigl( G_{b}^{ab} (\tau)\bigr)^{2}
G_{b}^{ab} (-\tau) \\
G_{b}^{aa} (\tau=0^{-} ) &= - S
\end{align}
\end{subequations}
Similarly for the fermionic model, we have :
\begin{subequations}\label{EqBasefermions}
\begin{align}
(G_{f}^{-1})^{ab} (i\omega_{n}) &=i\omega_{n}\delta_{ab} +
\lambda^{a}\delta_{ab} - \Sigma_{f}^{ab} (i\omega
_{n}) \\
\Sigma^{ab}_{f}(\tau ) &=\ -J^{2} \bigl( G_{f}^{ab} (\tau)\bigr)^{2}
G_{f}^{ab} (-\tau) \\
G_{f}^{aa} (\tau=0^{-} ) &=  q_{0}
\end{align}
\end{subequations}
In these equations, $\nu$ (resp. $\omega$) are the bosonic (resp.
fermionic)  Matsubara frequencies and the inversion should be
taken with respect to the replica indices $a,b$. Note that our
conventions for the sign of the Green functions in this paper
slightly differ from those  of Ref.~\onlinecite{SachdevYe}. Note
that, although the equations are written in term of $G$, the
physical quantity is the {\it local } spin susceptibility
$\chi_{loc} (\tau )=\moy{S (\tau )S (0)}$ which is given in the
large-$N$ limit by
\begin{equation}\label{DefChi}
\chi_{loc} (\tau )= G_{b}^{aa} (\tau)G_{b}^{aa} (-\tau )
\end{equation}
In many instances below (where we mostly focus on the bosonic case),
we shall drop the index $b$ in $G_b$.

\begin{figure}[ht]
\[
\fig{10cm}{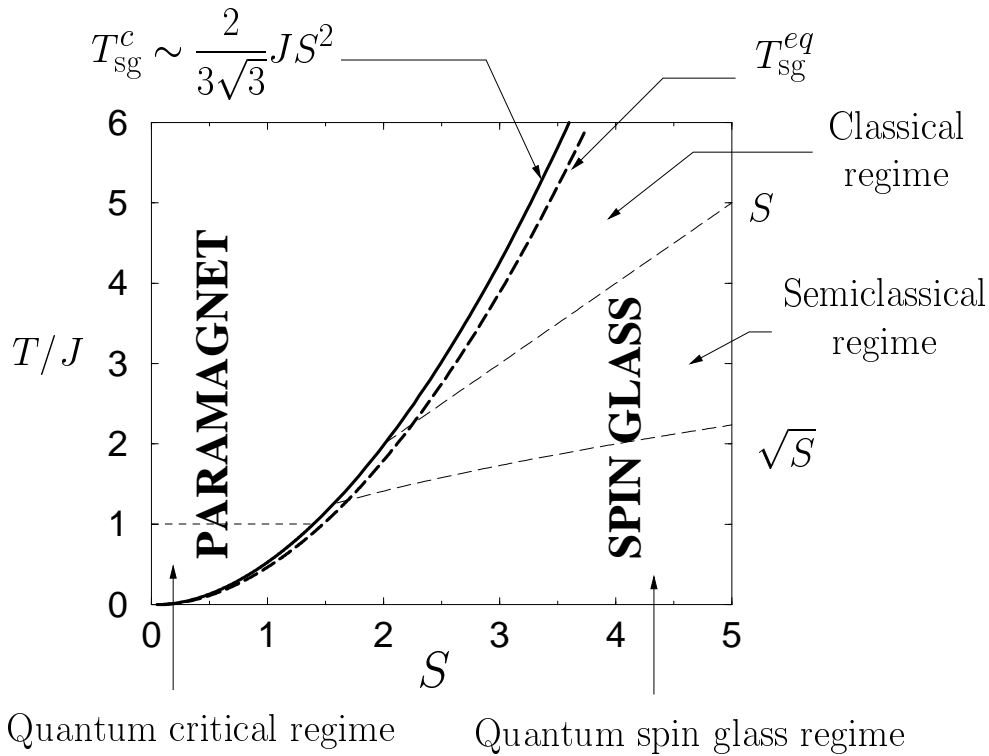}
\]
\caption{\label{DiagrammePhase}
\sl
Phase diagram of the mean field bosonic model. There is a spin glass
phase below the spin glass temperature $T_{sg}^{c}$, which is determined
with the marginality condition (See Section
\ref{SpinGlassPhases}). $T^{eq}_{sg}$ is the spin glass temperature as
determined with the stationarity criterion.}
\end{figure}

From both analytical and numerical analysis of these integral
equations, we have constructed the phase diagram displayed on Figure
\ref{DiagrammePhase}, as a function of the size of
the spin $S$ and the temperature $T$.
Let us give here a brief overview of the main features of this
phase diagram, which will be studied in great detail in the rest
of this paper. As is evident from Fig~\ref{DiagrammePhase}, it is
useful to divide the discussion into models with $S$ large and
$S$ small. Both regimes are accessible in the large $N$ limit,
where $S$ is effectively a continuous parameter taking all
positive values. For the physical case $N=2$, we will present
evidence later that at least $S=1/2$ is in the small $S$ regime
for the infinite-range model; moreover, we can expect that at least
some of the consequences of increased quantum fluctuations in
a realistic model with
finite-range interactions are mimicked by taking small $S$ values
in the large $N$ theory of the infinite-range model.

For large $S$, the ground state must clearly be a spin glass
(Fig~\ref{DiagrammePhase}). However, even for very large $S$, it is
necessary to consider quantum effects in understanding the low $T$
excitations and thermodynamics, and these have not been previously
described. In this paper (Sections~\ref{sec:largeS}
and~\ref{Thermo}) we will show that the local spin
susceptibility has a low energy density of states which increases
linearly with energy. At the same time, the specific heat also has
a linear dependence upon temperature. These results hold for
temperatures $T < J \sqrt{S}$, although characteristic excitations
have an energy of order $JS$ for $T < JS$; we will provide scaling
functions which determine the dynamic response functions at these
energies. At even higher $T$, there is a phase transition to a
paramagnet at $T \sim J S^2$. For large $S$ the static properties of this
phase transition
are well-described by a purely classical theory in which the
$\vec{S}$ in (\ref{DefSachdevYe}) are commuting vectors of length
$S$. Notice also that we indicate two critical temperatures,
$T_{sg}^c$ and $T_{sg}^{eq}$: as we discuss in Section
\ref{SpinGlassPhases}, these are a consequence of peculiarities in
the nature of replica symmetry breaking, where the dynamic
freezing into the spin glass phase ($T_{sg}^c$), happens
at a slightly higher temperature than the equilibrium transition
($T_{sg}^{eq}$).

For small $S$, we also find a spin glass phase
(Fig~\ref{DiagrammePhase}) at $T=0$, but the order vanishes at a
small $T$. Moreover, its excitations and finite $T$ properties
are very different from those at large $S$. These are now
dominated by signals of a ``spin-liquid'' state discussed in
Ref.~\onlinecite{SachdevYe} and described in
Section~\ref{SubSectSpinliquid}. In particular, we describe a
novel quantum-critical region in the paramagnet where
$\mbox{max}(\omega, T)$ is the characteristic energy scale, and
the local dynamic spin susceptibility obeys
$\chi^{\prime\prime}_{loc} (\omega ) \sim \tanh(\omega/2T)$. We
believe that aspects of this regime may be relevant to disordered
transition metal and rare earth compounds in regimes where
exchange interactions between the magnetic moments are playing a
dominant role. Completion of this picture requires an
understanding of the stability of the ``spin-liquid'' picture to
mobile charge carriers, and this has also been addressed in a
previous work \cite{Slush}.

\section{Paramagnetic phase}\label{ParamagneticPhases}

Contrary to the classical case, the paramagnetic phase of quantum
spin-glass models is non trivial in mean field theory. An early
discussion of these solutions has been given in
Ref.~\onlinecite{SachdevYe}, but we present here a much more
complete description, and compare our results to the $N=2$ case,
when numerical results are available. Since in this section we
look for paramagnetic solutions, we will consider only {\sl
replica diagonal} solutions of (\ref{EqBase}) : $G^{ab}\propto
\delta_{ab}$. Two types of paramagnetic solutions have been
found, that we will now consider successively : the {\it
spin-liquid} solutions and the {\it local moment} solutions.

\subsection{Spin-liquid solutions}\label{SubSectSpinliquid}

\subsubsection{The large-$N$ limit}

A low-frequency, long-time analysis of the integral equations
(\ref{EqBase}) reveals that, under the condition that
$\lambda-\Sigma(i0^+)$ vanishes at low temperature, a solution
can be found which displays a power law decay of the Green's
function at long time \cite{SachdevYe}: $G (\tau )\sim
1/\sqrt{\tau}$. These solutions display  a singularity in the
complex plane of frequencies, $z$, at $z =0$, with an amplitude
which can be parameterized by an angle $\theta$ as:
\begin{equation}\label{SingulariteSpinliquid}
G (z)\sim \frac{Ae^{-i\pi /4-i\theta }}{\sqrt{z}} \qquad \qquad
\text{ for } z\rightarrow 0,\  \Im z>0
\end{equation}
(Values of $z$ on the imaginary frequency axis at the Matsubara
frequencies will be denoted by $\nu_n$, while on the real axis
will be denoted, $\omega$.)
 Thus, these solutions display a slow local spin dynamics :
 $\Im \chi _{loc} (\omega )\propto \sgn \omega $ for
 $\omega \rightarrow 0, T=0$. Moreover, the {\it local} susceptibility
$\chi_{loc} (T)\equiv \int_{0}^{\beta }\chi_{loc} (\tau ) \, d\tau $
diverges as $\chi_{loc}
(T) \sim \ln T/J$ at low temperature.
More precisely, one can find the thermal scaling function characterizing
the $\tau \rightarrow \infty, T\rightarrow 0 $ limit, as explained in
a previous paper \cite{Slush} :
\begin{equation}\label{FormeEchelle}
\chi_{loc} (\tau ,\beta )
\propto \left(\frac{\pi /\beta }{\sin \pi \tau /\beta }
\right) + \dots \qquad \qquad \qquad
J\,\chi_{loc}^{\prime\prime}(\omega,T) \propto
\tanh\frac{\omega}{2T}
\end{equation}
Note that in the paramagnetic phase of  {\it quantum} models, the local
susceptibility $\chi_{loc} (T)$ (which is the response to a {\it
local} magnetic field) differs from the uniform susceptibility $\chi (T)$
(response to a constant magnetic field), contrary to {\it classical}
spin glass models where $\chi = \chi_{loc}$ \cite{MezardParisiBook} :
this is a consequence of
the commutation relations of the spin, as can be seen for example in
the high-temperature expansion in the $SU (2)$ model.
In this large-$N$ limit, it can be  shown that $\chi (T) \ll
\chi_{loc} (T)$ for $T\rightarrow 0$ and numerical computations
indeed suggest that $\chi (T)\sim \mbox{const.}$ \cite{Slush}.

Remarkably, the parameter $\theta$ which characterizes the {\sl
spectral asymmetry} \cite{KondoLong} of the spectral density at
low frequency can be explicitly related to the size $S$ of the
spin (which involves a priori an integral of the spectral density
over all frequencies). This is very similar to a kind of Friedel
sum rule applying to this problem, and indeed the derivation
follows a very similar route, based on the existence of a
Luttinger-Ward functional. (Interestingly enough, the ``boundary
term'' which usually vanishes in such derivations contributes
here a finite value). This derivation is presented in detail in
Appendix \ref{AppSpectralAsym}, where the following relation
between $\theta$ and $S$ is established:
\begin{equation}\label{ValueOfTheta}
\frac{\theta }{\pi } + \frac{\sin 2\theta }{4} =
\left\{
\begin{aligned}
\frac{1}{2} + S& \text{ \qquad in the bosonic model}\\
\frac{1}{2} - q_{0}& \text{\qquad  in the fermionic model}
\end{aligned}
\right.
\end{equation}
This relation has important consequences for the physical properties of the
spin-liquid solutions. First, we note that the spectral density must
obey the positivity conditions: $\Im G_f(\omega+i0^+) < 0$ and
$\mbox{sgn}(\omega)\Im G_b(\omega+i0^+) < 0$.
Hence, in the fermionic case, $\theta$ must obey $-\frac{\pi }{4}\leq \theta \leq
\frac{\pi }{4}$. It is easily checked from (\ref{ValueOfTheta}) that
$\theta$ precisely describes this range of parameters as $q_0$ is varied
from $q_0=0$ to $q_0=1$, and that the $\theta(q_0)$ relation is unique.
This suggests that the spin-liquid solution
is an acceptable low-temperature solutions for the whole range of
$q_0$ in the fermionic case.
\begin{figure}[ht]
\[
\fig{6cm}{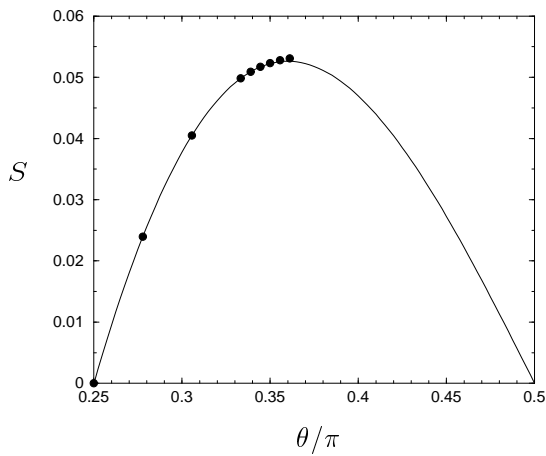}
\]
\caption{\label{PlotOfTheta}
\sl
$S$ as a function of $\theta$ for the bosonic model.
 The solid line is given by relation
(\ref{ValueOfTheta}) and the points were obtained previously from a
numerical solution of the saddle-point equation at zero
temperature \cite{SachdevYe}.}
\end{figure}
In contrast, in the bosonic case, the plot in Fig.\ref{PlotOfTheta}
shows that (\ref{ValueOfTheta}) actually defines {\it two} values of
$\theta$ (in the allowed range $\frac{\pi }{4}\leq \theta \leq
\frac{3\pi }{4}$) for a given spin $S$ as long as $S<S_{max}\simeq
0.052$, while no value of $\theta$ is found for $S>S_{max}$. This
implies that no paramagnetic solution of the spin-liquid type is found
{\it at zero-temperature} in the bosonic case as soon as $S>S_{max}$
(note that furthermore $S_{max}$ is very small). For $S<S_{max}$, such
solutions exist at zero temperature
 (even though they are not the true ground-state,
see below) with the locally stable solution corresponding to the
smallest of the two values of $\theta$.  However, even for $S>S_{max}$,
{\it at low ($T<J$) but finite temperature}, the spin liquid solutions
do exist in the bosonic model.  By this, we mean that a
numerical computation in imaginary time gives a solution which
exhibits the scaling form (\ref{FormeEchelle}), for which a
unambiguous value of
the spectral asymmetry $\theta$ can be defined and computed
numerically.
At very low temperature, these solutions are unstable to the spin glass solution,
but above the spin glass temperature at low spin, they are relevant
 in the quantum critical regime associated
with the quantum critical point at $S=0$.
We shall comment in more detail, at the end of the following section, on the
nature of the paramagnetic solutions found at low temperature for small
values of $S$, in the bosonic case.

\begin{figure}[ht]
\[
\fig{6cm}{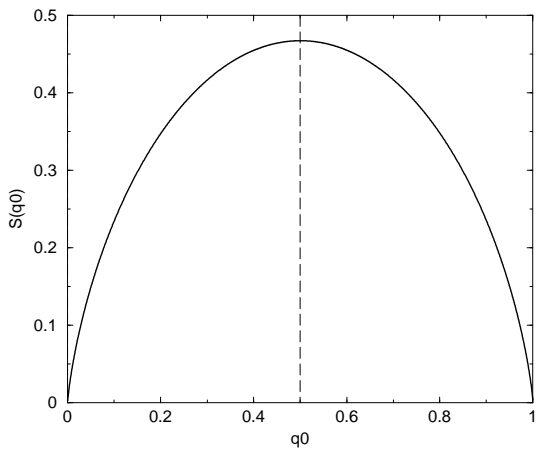}
\]
\caption{\label{PlotEntropy}
\sl
Entropy as a function of the size of the spin ($q_{0}$) in the
fermionic model.}
\end{figure}

Another  consequence of relation
(\ref{ValueOfTheta}) is that it allows to predict that these
spin-liquid solutions have a {\it non-zero extensive entropy} at
zero-temperature and to calculate the value of this entropy
analytically. The derivation of this result follows very closely a
similar analysis of the overscreened multichannel Kondo problem
in the large-N limit, performed in
Ref.~\onlinecite{KondoLong,KondoPRL} and only the main steps will
be repeated here. This can be done either in the bosonic model or
in the fermionic one, with slight modifications. Since the spin liquid
solutions are relevant at zero temperature only in the fermionic model,
we shall present the result in this case.
First, denoting by ${\cal S}$ the value of the
entropy per spin at zero temperature, one establishes the
following thermodynamic equality:
\begin{equation}
\frac{\partial{\cal S}}{\partial q_{0}} =
-\frac{{\partial\lambda}}{{\partial T}}|_{T=0}
\end{equation}
Then, a low temperature expansion is used which allows to relate the slope
of $\lambda(T)$ to
the spectral asymmetry parameter $\theta$ above, so that one
finally gets (in the fermionic case):
\begin{equation}
\frac{\partial{\cal S}}{\partial q_{0}} =
 \ln
\frac{\sin(\pi/4-\theta)}{\sin(\theta+\pi/4)}
\end{equation}
The entropy is then obtained by integration over the size of the
spin, with the physically obvious boundary conditions
${\cal S}(q_0=0)={\cal S}(q_0=1)=0$. The resulting
value of the entropy as a function of $q_0$ is plotted in
Fig.\ref{PlotEntropy}.

Finally, we comment on the physical nature of the spin-liquid
paramagnetic solutions found in this section. These solutions
correspond to a partial screening of the local moment at each
site, due to the interaction with the other spins. As a result
the local susceptibility diverges logarithmically (much slower
than a Curie law), but an extensive entropy is still present at
$T=0$, indicating a degenerate state. From a local point of view,
the physics is somewhat similar to an overscreened Kondo system,
but here the gapless bath which quenches the spin is not external
but self-consistently generated by the other spins. We suspect
that the physics of this phase has to do with the degeneracy of
the (large-$N$ generalization) of the ``triplet'' state in which two
spins are bound whenever a strong ferromagnetic bond $J_{ij}$ is
encountered. In Section \ref{SpinLiquidN=2}, we show that this
spin-liquid regime is not a peculiarity of the large-$N$ limit but
indeed survives in the mean-field description of the quantum
critical regime of a $SU(2)$ quantum Heisenberg spin-glass. It
would be very valuable to gain a more direct understanding of
this gapless spin-liquid regime from a study of the problem for a
fixed configuration of bonds, before averaging over disorder.
This could be achieved numerically and is left for future studies.

\subsubsection{Beyond the large-$N$ limit}
\label{SpinLiquidN=2}
This subsection will show how
recent renormalization group analysis of related models\cite{AMS,SBV,VBS}
imply that
the above spin-liquid solution applies to all orders in $1/N$. In
particular, the large $N$ solution with $\mbox{Im} \chi_{loc}
\propto \mbox{sgn}~ \omega$ for small $\omega$ acquires no
corrections to its functional form: the only changes are to the
non-universal proportionality constant.
All the discussion below will be in a paramagnetic
phase where it is sufficient to consider only a single replica,
and so we will drop replica indices in this subsection.

We begin by rewriting (\ref{ActionEffective}) in the following
form\cite{AMS}
\begin{equation}
S_{eff} = S_B - \gamma_0 \int_0^{\beta} d\tau \overrightarrow{S} (\tau)
\cdot \overrightarrow{\phi} (\tau)
\label{ss1}
\end{equation}
where $\gamma_0$ is a coupling constant and
$\overrightarrow{\phi}$ is an annealed Gaussian random field
with $\langle \overrightarrow{\phi} (\tau) \cdot
\overrightarrow{\phi} (0) \rangle = 1/|\tau|^{2-\epsilon}$.
It is reasonable to expect that the spin correlations in the
quantum ensemble defined by (\ref{ss1}) decay with the power-law
$\langle  \overrightarrow{S} (\tau) \cdot
\overrightarrow{S} (0) \rangle \sim 1/|\tau|^{\sigma}$, and we are
interested in determining the value of the exponent $\sigma$.
A simple extension\cite{SachdevYe} of the
solution discussed above implies that in the large $N$ limit
$\sigma = \epsilon$. Here we will argue that this equality is in fact
 {\em exact for all $N$}. Now using the self-consistency condition
(\ref{CondCoherence}) we obtain $\sigma=\epsilon=1$, which then
implies $\mbox{Im} \chi_{loc}
\propto \mbox{sgn}~ \omega$.

The field-theoretic renormalization group analysis of (\ref{ss1})
was discussed in Ref.~\onlinecite{VBS}, and we will highlight the
main results. The key observation is that renormalization of the
theory (\ref{ss1}) requires only a single wave-function
renormalization factor $Z$, and that there is no independent
renormalization of the coupling constant $\gamma_0$. This result
was established diagrammatically in Ref.~\onlinecite{VBS}, and we
will not reproduce the argument here. So if we renormalize the
spin by $\overrightarrow{S} = \sqrt{Z} \overrightarrow{S}_R$,
then the coupling constant renormalization is simply $\gamma_0 =
\mu^{\epsilon/2} \gamma /\sqrt{Z}$, where $\mu$ is a
renormalization scale. The renormalization constant is in general
a complicated function of $\gamma$, and was determined to
two-loop order in Ref.~\onlinecite{VBS}:
\begin{equation}
Z = 1 - \frac{2\gamma^2}{\epsilon} + \frac{\gamma^4}{\epsilon} +
\ldots
\label{ss2}
\end{equation}
in a minimal subtraction scheme. However, even though $Z$ is not
known exactly, the exponent $\sigma$ can be determined exactly.
Standard field-theoretical technology shows that the above
renormalizations imply the $\beta$-function
\begin{equation}
\beta (\gamma) = - \frac{\epsilon \gamma}{2} \left( 1 -
\frac{1}{2} \frac{\partial \ln Z}{\partial \ln \gamma}
\right)^{-1}.
\label{ss3}
\end{equation}
Furthermore, the exponent $\sigma$ is given by the value of
\begin{equation}
\sigma (\gamma) = \beta(\gamma) \frac{\partial \ln Z}{\partial
\gamma}
\label{ss4}
\end{equation}
at the fixed point $\gamma=\gamma^{\ast}$ where $\beta(\gamma)$
vanishes. Comparing (\ref{ss3}) and (\ref{ss4}) we see that
\begin{equation}
\beta (\gamma) = -(\epsilon - \sigma(\gamma)) \gamma/2.
\label{ss5}
\end{equation}
Clearly, a zero of the $\beta$ function must have $\sigma =
\epsilon$, and this establishes the required result.

We note that similar examples of a critical exponent being valid to all
orders (in spite of a non-trivial $\beta$-function) can be found for
other models in the statistical mechanics of disordered systems
(see e.g \cite{forster,honko}).

\subsection{``Local moment'' solutions}

In a mean field  model, one usually expects to find locally stable
(while possibly unstable to ordering) paramagnetic solutions of the
mean-field equations down to zero temperature. Hence, the absence
of solutions of the spin-liquid type for $S>S_{max}$ suggests that
a different kind of paramagnetic solution should exist for those
values of the spin. Indeed, we have found that the integral equations
(\ref{EqBase}) have another class of paramagnetic solutions
in the bosonic case.
These solutions actually exist for all values of the spin $S$ and down
to zero temperature. Hence, they coexist at low $S$ with the spin-liquid
solutions in some range of temperature.
Their physical nature is very different from the previous spin-liquid
solutions, and as discussed below they are not very physical solutions
when considered at low temperature. They are characterized
by a Green's function which does not decay at long times and obeys the
asymptotic behavior:
$G_{b} (\tau ) \simeq  -S - e^{-J^{2}\beta S^{3}\tau}$. In contrast to
the spin-liquid case, $\lambda$ diverges for $T\rightarrow 0$ in this
regime:
\begin{equation}\label{NewParaLambdaLowT}
\lambda \sim \frac{J^{2} S^{3}}{T}
\end{equation}
Finding numerically these solutions of (\ref{EqBase}) requires
some care. We have used an algorithm in which
we solve (\ref{EqBase}) in imaginary time for $G(\tau)$, for a fixed
value of $r=\int_{0}^{\beta }G (\tau )d\tau$,
and then adjust the number of particles to $S$ by a dichotomy on $r$.
\begin{figure}[ht]
\[
\fig{6cm}{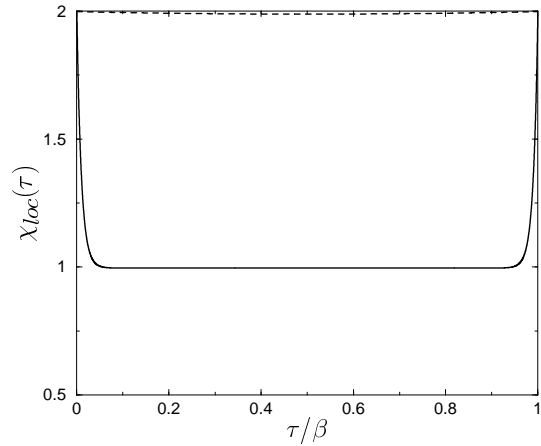}
\]
\caption{\label{KiNewParaFig}
\sl
$\chi_{loc} (\tau )$ extracted from a numerical solution
of saddle point equation (\ref{EqBase}) in
imaginary time, for $S=1$, $J=1$.
The solid curve is low temperature ($J\beta =10$), the dashed curve is
high temperature ($J\beta =.1$); for intermediate temperatures, the
curves interpolate between the two.}
\end{figure}
The local susceptibility $\chi_{loc} (\tau )$ obtained in this manner
is displayed on Figure
\ref{KiNewParaFig}. At high temperature, we find $\chi_{loc} (T)=
\frac{S (S+1)}{T}$ as expected since the spin is
essentially free. At low temperature, we find  another Curie law, with a
reduction of the  Curie constant due to quantum fluctuations  :
\begin{equation}\label{ReducedCurie}
\chi_{loc} (T) = \frac{S^{2}}{T} \qquad\qquad  \hbox{  for  }
T\rightarrow 0
\end{equation}
Hence, for these solutions, the effect of the interactions with the other
spins is not strong enough to result in a qualitatively different
screening regime, resulting merely in a reduction of the Curie constant.
This is analogous to an {\it underscreened} Kondo regime.

\begin{widetext}
\begin{figure}[ht]
\[
\includegraphics[height=10cm]{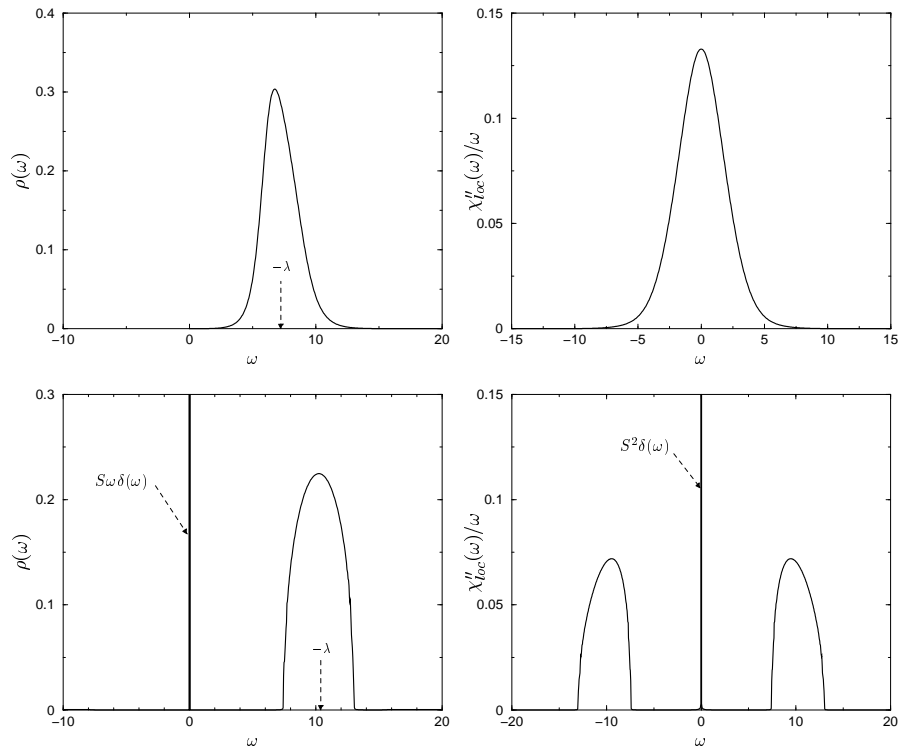}
\]
\caption{
\sl
 \label{NewParaOmega}
Spectral densities for $G_{b}$ (left) and for the local
susceptibility $\chi_{loc}$ (right), for high temperature (top)  and a
very low temperature (bottom).  These results are
extracted from a numerical solution of saddle point equation (\ref{EqBase}) in
real frequencies.}
\end{figure}
\end{widetext}

These solutions are of a similar type than those found in in a
quantum Monte-Carlo simulation of the $SU (2)$ model
\cite{Grempel1}. There also, a reduction of the Curie constant
from $S (S+1)/3$ to $S^{2}/3$ was clearly observed. Thus,
contrary to one of the conclusions of Ref.~\onlinecite{Grempel1},
the large-$N$ limit correctly reproduces the paramagnetic local
moment solution found for the physical $N=2$ case. Moreover, a
numerical solution of the large-N integral equations
(\ref{EqBase}) for real frequencies can also be obtained both at
high and low temperature. The results are presented in Figure
\ref{NewParaOmega} for both $\rho_{b} (\omega )\equiv -1/\pi  \Im
G_{b} (\omega )$ and $\chi_{loc}'' (\omega )\equiv 1/\pi \Im
\chi_{loc} (\omega )$. At high temperature, $\rho_{b}$ is centered
around $\lambda \sim T\ln ((S+1)/S)$ and $\chi_{loc}'' (\omega )$
is a simple peak. At low temperature, we find in  $\chi_{loc}''
(\omega)$ a delta-function peak at zero frequency with weight
$S^{2}$, which is associated with longitudinal relaxation (see
Ref.~\onlinecite{Grempel1} for a discussion for $N=2$)  and two
peaks, centered around $\pm \lambda \propto 1/T$ (Cf Eq.
(\ref{NewParaLambdaLowT})) with a constant width, which are
associated with transverse relaxation. Again, these results are
very similar to the conclusions reached in
Ref.~\onlinecite{Grempel1} from a fit of the imaginary-time data
for $N=2$, using a model $\chi^{''}(\omega)$. The only difference
is that the central peak is not broadened by thermal fluctuations
in the large-$N$ limit.
Finally, since these solutions describe the formation of a
local moment at low
temperature, and that the onset of the quenching temperature
(at which the reduction of the Curie constant sets in) turns out to be
lower than the temperature where spin-glass ordering occurs (as shown in the
next section),
we consider these solutions as a mean-field artefact of the spin glass
ordering.  Static limits of such solutions are actually known
to occur in the classical SK model. We also note that the internal
energy of these solutions have an unphysical divergence
as $T\rightarrow 0$.

We close this section by noting that, for small values of the spin
$S$, a rather intricate pattern emerges for the stability and
coexistence of the two kinds of paramagnetic solutions described
above. We have studied this numerically in some detail but do not
report this here, since most of these phenomena occur below the
spin-glass ordering temperature anyway. The important features
have been displayed on Fig.\ref{DiagrammePhase}. We emphasize
again that the spin-liquid solutions are the relevant solutions
describing the whole quantum-critical regime. Even though they
are unstable at $T=0$ for $S>S_{max}$ they remain consistent
{\it finite temperature} solutions in a much wider range of values
of $S$ for $T<J$. At higher $S$ and above the spin-glass ordering
temperature, the paramagnetic phase behaves as a local moment,
with a Curie constant getting gradually reduced as $T$ is lowered.

\section{The spin-glass phase}\label{SpinGlassPhases}

In this section, we investigate spin-glass ordering in this model.
The first observation that we make (Sec.\ref{sg_susc}) is that the
spin-glass susceptibility (i.e the response to a spin-glass ordering field)
is actually of order $1/N$ in the large-N limit. This does not
preclude a spin-glass phase, but means that the transition is not
associated associated with a linear instability. Indeed, we shall
find explicit solutions in the ordered phase in the bosonic case,
while the fermionic case does not have a spin-glass phase at
$N=\infty$ (but does order as soon as $1/N$ corrections are
considered).

\subsection{Spin-glass susceptibility}\label{sg_susc}

Here, we derive an exact expression for the spin-glass
susceptibility, valid for arbitrary $N$ in this mean-field model.
The derivation is most conveniently performed using replicas. We
note that in the presence of spin-glass order, the correlation
function $\moy{\overrightarrow{S}^{a} (\tau )\cdot
\overrightarrow{S}^{b}(\tau')}$ acquires a non-zero value for
$a\neq b$, which is however static, i.e independent of
$\tau-\tau'$ (see e.g Ref.~\onlinecite{SachdevBook}). This crucial
point is due to the fact that different replicas are independent
of one another before averaging, so that for $a\neq b$:
\begin{equation}
\overline{\moy{\overrightarrow{S}^{a}(\tau )\cdot
\overrightarrow{S}^{b}(\tau')}} =
\overline{\moy{\overrightarrow{S}^{a} (\tau )}\cdot
\moy{\overrightarrow{S}^{b}(\tau')}} =
\overline{\moy{\overrightarrow{S}^{a} (0)}\cdot
\moy{\overrightarrow{S}^{b}(0)}}
\end{equation}
In the following, we shall denote by $q_{ab}$ the (normalized) off-diagonal
correlation function which is an order parameter for the
spin-glass phase:
\begin{equation}
q_{ab} \equiv \frac{1}{N^2} \sum_{a\neq b}
\overline{\moy{\overrightarrow{S}^{a} (\tau )\cdot
\overrightarrow{S}^{b}(\tau')}}
\end{equation}
We consider the stability of the paramagnetic phase to this type
of ordering, and introduce an ordering field $H_{ab}$ conjugate to
$q_{ab}$. This has two effects:

i) It adds to the effective action (\ref{ActionEffective}) an
explicit term:
\begin{equation}
\delta S = \frac{1}{N}
\int d\tau\int d\tau' \sum_{a\neq b} H_{ab}\,
\vec{S}_a(\tau)\cdot\vec{S}_b(\tau')
\end{equation}
The normalization of $H_{ab}$ has been chosen in such a way
that the change in the total free-energy is of order $N$.

ii) It modifies the value of the self-consistent field $Q^{ab}$.
The change of the off-diagonal component, to linear order, is
imposed by the self-consistency condition (\ref{CondCoherence})
to be: $\delta Q^{ab}=\delta q_{ab}$, with $\delta q_{ab}$ the
induced order parameter.

One can then perform an expansion of the off-diagonal correlation
function up to linear order in both $H_{ab}$ and $\delta Q^{ab}$.
This yields:
\begin{equation}
\delta q_{ab} = \frac{1}{N}
\chi_{loc}^2\,\left(H_{ab}+J^2\delta Q^{ab}\right)
\end{equation}
in which $\chi_{loc}$ is the local susceptibility of the
paramagnetic phase. Since $\delta Q^{ab}=\delta q_{ab}$, this
finally yields the susceptibility to spin-glass ordering:
\begin{equation}
\label{ChiSG}
\chi_{sg} \equiv \frac{\delta q_{ab}}{H_{ab}} =
\frac{1}{N}\,\frac{\chi_{loc}^2}{1-(J\chi_{loc})^2/N}
\end{equation}
This formula has two important consequences. The susceptibility
to spin-glass ordering is of order $1/N$
and any spin-glass instability at $N=\infty$ must be associated
with a non-linear effect of higher order (i.e come from terms of
higher than quadratic order in $q_{ab}$ in the free energy).
Furthermore, (\ref{ChiSG}) shows that for finite $N$, a (linear)
instability into a spin-glass phase will occur when
$J\chi_{loc}(T)=\sqrt{N}$. Hence the fermionic model, for which
$\chi_{loc}$ diverges at low-T, will have a spin-glass instability
for arbitrary large but finite $N$. More precisely, since
the low-temperature behavior $J\chi_{loc} \sim \ln \frac{J}{T}$
has been shown above to hold for all N for the paramagnetic solution of
the fermionic case, we conclude from (\ref{ChiSG}) that the spin-glass
transition temperature depends on $N$ in that case as :
$T_c^{f}\sim J e^{-\sqrt{N}}$.

\subsection{Spin glass solutions and the ``replicon'' problem}

We now turn to the explicit construction of solutions of the
integral equations (\ref{EqBase}) with spin-glass ordering, in the
bosonic case. The same reasoning as above shows that the Green's
function $G^{ab} (\tau )$ does not depend on $\tau$ for $a\neq b$.
Thus the most general Ansatz for the Green's function $G^{ab}$ can be
written as :
\begin{equation}\label{AnsatzParisi}
G^{ab} (\tau ) =\left\{
\begin{aligned}
\widetilde{G} (\tau ) - g_{1}   & \qquad \qquad (a=b)\\
 - g_{ab}  & \qquad \qquad (a\neq b)\\
\end{aligned}
\right.
\end{equation}
where $\widetilde{G} (\tau )$ is a function of the imaginary time,
$g_{ab}$ a constant matrix and $g_{1}$ a constant.
By definition, $g_{1}$ is fixed  so that  $\widetilde{G}$ is
regular at $T=0$, {\sl i.e. } $\widetilde{G} (\tau) \rightarrow 0$ as
$\tau \rightarrow \infty$.
In the following discussion, we will restrict ourselves to solutions
given by a Parisi Ansatz for $g_{ab}$ (a replica symmetry breaking scheme).
In the $n\rightarrow 0$ limit (where $n$ is the number of replica),
this matrix becomes a function $g (u)$ of a
continuous variable $u$ with  $0\leq u\leq 1$ \cite{MezardParisiBook}.

Our equations (\ref{EqBase}) involve the Green function $G$, whereas the
physical quantity is the susceptibility $\chi^{ab} (\tau )=G^{ab}
(\tau )G^{ab} (-\tau )$. The order parameter $q_{ab}$ widely introduced in
the spin glass literature \cite{MezardParisiBook} is
given here by $q_{ab} =g_{ab}^{2}$, or in the limit  $n\rightarrow 0$ :
\begin{equation}
q (u) = g (u)^{2}
\end{equation}
The Edwards-Anderson parameter is  $q_{EA}  = q (1)=g (1)^{2}$
\cite{MezardParisiBook,SachdevBook}. Since at zero temperature,
in the long time limit, we find $\lim_{\tau \rightarrow \infty
}{G^{aa} (\tau )G^{aa} (-\tau )} = q_{EA}$ we find $g_{1} = g
(1)$, by definition of $g_{1}$. More precisely, we look for
solutions in which these two definitions of $q_{EA}$ coincide but
it is not really an assumption in our computation : if this
relation was violated, we would simply find for $\widetilde{G}$ a
non vanishing limit for $\tau \rightarrow \infty$.

Among the various possible replica symmetry breaking schemes
\cite{MezardParisiBook}, we will now focus on {\it one-step}
solutions, since we have not found any other,  either two-steps
or with continuous replica symmetry breaking. In this case, the
function $g(u)$ is piecewise constant : $g (u)= \overline{g}$ for
$0<u<x$ and $g(u)=g$ for $x\leq u\leq 1$. In the following, we
will refer to $x$ as the {\sl breakpoint}. According to
(\ref{EqBase}), the self-energy $\Sigma$ has the form :
\begin{equation}\label{AnsatzParisiSigma}
\Sigma^{ab} (\tau ) = \left\{
\begin{aligned}
\widetilde{\Sigma} (\tau ) - J^{2} g^{3}   & \qquad \qquad (a=b)\\
                     -J^{2} g_{ab}^{3}   & \qquad \qquad (a\neq b)\\
\end{aligned}
\right.
\end{equation}
with $\widetilde{\Sigma }$ is given by (\ref{EqSG_DefSigTilde}).
Using the standard formulas to invert the Parisi matrices
in the limit $n\rightarrow 0$ \cite{MezardParisiManifold}, we find :
\begin{align}\label{Intermed1}
 \widetilde{G}^{-1} (i\nu_{n}) = i\nu_{n} + \lambda
-\widetilde{\Sigma}(i\nu_{n})  \\
\label{Intermed2} J^{2} g^{2} \widetilde{G}(i\nu_n = 0)^{2} = 1 +
J^{2} \beta x g^{3} \widetilde{G} (i \nu_n = 0)
\end{align}
At this stage, it is useful to introduce here a new parameter $\Theta$
defined by :
\begin{equation}\label{DefTheta}
\widetilde{G} (i \nu_n  ) = -\frac{\Theta}{Jg}
\end{equation}
We then eliminate $\lambda$ in (\ref{Intermed1}) and $\widetilde{G} (0)$
in (\ref{Intermed2}) and obtain finally a closed set of equations for
$\widetilde{G}$ and $g$ :
\begin{subequations}\label{EqSG}
\begin{equation}\label{EqSG_Dyson}
\left( \widetilde{G} (i\nu_{n})\right)^{-1} = i\nu_{n} - \frac{Jg}{\Theta } -
\left(\widetilde{\Sigma} (i\nu_{n}) - \widetilde{\Sigma} (0) \right)
\end{equation}
\begin{equation}\label{EqSG_DefSigTilde}
\widetilde{\Sigma} (\tau ) = J^{2} \left(
\widetilde{G}^{2} (\tau )\widetilde{G} (-\tau ) -
 2 g \widetilde{G} (\tau ) \widetilde{G} (-\tau)  -  g
\widetilde{G}^{2} (\tau )  + 2 g^{2} \widetilde{G} (\tau )
+  g^{2} \widetilde{G} (-\tau )\right)
\end{equation}
\begin{equation}\label{EqSG_NumberParticle}
\widetilde{G} (\tau =0^{-}) = - (S - g)
\end{equation}
\begin{equation}\label{EqSG_x}
\beta x = \frac{1}{J g^{2}} \left(\frac{1}{\Theta} - \Theta  \right)
\end{equation}
\end{subequations}
The crucial observation at this point is that  these
saddle-point equations possess a one parameter family of solutions,
parametrised by  $\Theta $ or equivalently by the breakpoint $x$.
This phenomenon already occur in other models which have
a {\sl one-step} replica symmetry breaking solution \cite{GiamLedou}.
The determination of the breakpoint turns out to be the most difficult
question of this analysis.
Two possible criteria are :
\begin{enumerate}
\item To minimize the free energy ${\mathcal F}(x)$ as a function of $x$, as would be
required by the thermodynamics. In the
following, we will refer to this as the  {\sl equilibrium
criterion}. This criterion has been used in a previous attempt to
understand this spin glass phase \cite{kopec}.
\item To impose a vanishing lowest eigenvalue of the fluctuation matrix
in the replica space. We will refer to this as
the {\sl marginality} or {\sl replicon } criterion.
Although it is not really justified up to now, we
will argue that it is the correct choice.
\end{enumerate}
This problem is not due to  the quantum aspect of our model :
it already appears similarly in some classical spin glass
models, in the $p$-spin model for example.
In this classical model, the study of the dynamics shows the existence
of a dynamical transition  $T^{dyn}$ above the static spin glass
temperature $T^{eq}$ given by the  static solution of the mean field
model. It turns out that, in this classical  model, the
replicon criterion has been proven to give
the same transition temperature $T^{c}=T^{dyn}$.
Moreover, it has been shown \cite{CugliandoloPspinQ,LeticiaPspin}
that the same
phenomenon occurs in some {\sl quantum } version of the p-spin model.
Thus, using this condition, it is possible in some sense to mimic the
dynamics by simply solving a static problem, although this is not
fully understood at present.

In the present model, the two criteria give a coherent solution
but with totally different spectra of {\it equilibrium}
fluctuations : the equilibrium criterion leads to a gap in $\chi''
(\omega )$ whereas the replicon criterion is the only one which
give a {\sl gapless } $\chi''(\omega )$ (a similar observation was
made in Ref.~\onlinecite{GiamLedou} in a one dimensional quantum
model with disorder). We believe that in this quantum Heisenberg
spin glass the replicon criterion  provides us with the correct
physical solution ($T_{c}$), contrary to the equilibrium
solution, which gives the static transition temperature
($T_{eq}$). However this claim cannot be proved in the present
context : in particular, the static solution does give a full
solution of (\ref{EqBase}). A study of the true Hamiltonian
dynamics in real time and finite temperature of this quantum
problem is necessary for a deeper understanding of this question,
but this is beyond the scope of this paper. Let us now examine
the two criteria separately in more details.

\subsubsection{The replicon criterion}
\label{sec:replicon}
To apply the replicon criterion, we need to study the
fluctuations of the free energy in the replica space around the one-step
solution.
In the large-$N$ limit, the free energy is given by the expression :
\begin{equation}\label{FormuleF1}
{\mathcal F}[G^{ab}, \lambda ] = \frac{1}{\beta } \sum_{n} \Tr
\ln \Bigl(i\nu_{n} + \lambda  - \Sigma^{ab} (i\nu_{n}) \Bigr) +
\frac{3J^{2}}{4} \sum_{ab} \int_{0}^{\beta} d\tau \, \left[
G^{ab} (\tau ) G^{ab} (-\tau) \right]^2  - \lambda S
\end{equation}
Under infinitesimal variation   $\delta g_{ab}$ for $a\neq b$,
 the variation of the free energy is (up to second order)
\begin{equation}\label{DefMatrixFluctuations}
\delta {\mathcal F}=\sum_{a>b\atop c>d} M_{ab,cd}\delta
g_{ab}\delta g_{cd}.
\end{equation}
Strictly speaking, as this is a quantum problem,
we have to simultaneously consider the
variation of the diagonal component, $\delta\widetilde{G}(\tau)$ in
(\ref{AnsatzParisi}) in a study of the fluctuations. In a spin
glass phase, there is indeed a coupling between $\delta g_{ab}$
and $\delta \widetilde{G} (\tau)$ which modifies the fluctuation
eigenvalues. Fortunately however, as we show in Appendix
\ref{AppReplicon}, this coupling does not modify the eigenvalue
$e_{1}$ and our main result (\ref{ThetaMarginality}) below, and so
we will neglect $\delta \widetilde{G} (\tau)$ here.
The diagonalization of the $n (n-1)/2 \times n (n-1)/2 $ matrix $M$ is
briefly explained in Appendix \ref{AppReplicon} and gives three eigenvalues
\begin{eqnarray}\label{EigenvaluesFluct}
\nonumber
&e_{1} = 3\beta  J^{2}g^{2} (1-3\Theta^{2})\\
&e_{2} = \frac{3\beta  J^{2}g^{2}}{\Theta^{2}}\left(\Theta^{2} -3
+3\beta Jg^{2}\Theta (1+\Theta) \right)\\
\nonumber
&e_{3} = 6\beta  J^{2}g^{2}\left(3\beta  J g^{2} \Theta -1 \right)
\end{eqnarray}

A first consequence of this analysis is that replica symmetric
solutions are unstable, since from Eq. (\ref{EqSG_x}) they correspond
to $\theta =1$ and then $e_{1}<0$.
Hence, these solutions will not be considered in the following discussion.

A full solution of Eqs.(\ref{EigenvaluesFluct}) is required to show the
positivity of $e_{2},e_{3}$, but we immediately see that $e_{1}=0$ for
\begin{equation}\label{ThetaMarginality}
\Theta_{R}= \frac{1}{\sqrt{3}}
\end{equation}

Quite remarkably, we will see below in Section~\ref{sec:largeS}
and Appendix~\ref{app:gapless} that precisely the same
value of $\Theta$ is selected by a criterion which is seemingly
entirely independent. We will study the dynamic spectral functions
in the spin-glass phase, as defined by $\widetilde{G}(\tau)$,
and show that their associated spectral densities are non-zero
as $|\omega| \rightarrow 0$ only for the value of $\Theta$ in
(\ref{ThetaMarginality}). So marginal stability in replica
space appears to be connected to a gapless quantum excitation
spectrum. We may intuitively understand this as due to the
availability of many low energy states when the system first
freezes, but a better understand should emerge from a real-time
analysis.

\subsubsection{The equilibrium criterion}
To apply the equilibrium criterion, we start from the expression
of the free energy ${\mathcal F}$ and solve for $x$ :
\begin{equation}\label{Critere1}
\frac{d{\mathcal F} (x)}{dx}=0
\end{equation}
The computation of the {\sl total} derivative (\ref{Critere1})
reduces to $\partial_{x}{\mathcal F} (x)|_{\widetilde{G},\lambda}$
because the saddle-point equations (\ref{EqBase}) {\sl for finite
$n$} are equivalent to
\begin{equation}\label{PropF1}
\frac{\partial {\mathcal F}}{\partial G^{ab}} =\frac{\partial
F}{\partial\lambda }= 0
\end{equation}
as can be checked by an explicit calculation. Computing the
logarithm in (\ref{FormuleF1}) (using Appendix II of
Ref.~\onlinecite{MezardParisiManifold}) and taking the derivative
leads to :
\begin{equation}
\frac{3}{4} J^{2} g^{4} - \frac{2}{(\beta x)^{2}}  \ln \bigl(-J g
\widetilde{G} (i\nu_n = 0)) = - \frac{ g}{\beta x \widetilde{G}
(i\nu_n = 0) }
\end{equation}
and finally to a equation for  $\Theta$ :
\begin{equation}\label{ThetaStationnarity}
2 \ln  \Theta  + \frac{1}{4\Theta^{2}} + \frac{1}{2} -
\frac{3\Theta^{2} }{4}=0
\end{equation}
This equation has two solutions : the replica symmetric one
$\Theta=1$ (unstable, as explained above),
and a non trivial one $\Theta =\Theta_{eq}\approx 0.4421\dots$.
Contrary to the previous solution, we will
see in Section~\ref{sec:largeS} that $\Im \widetilde{G}$ has a gap
for this value of $\Theta$.

\subsection{The phase diagram}

Once $\Theta $ has been determined, the equations (\ref{EqSG}) can be
solved either numerically (both in
imaginary time and in real frequency) or analytically in the
$S\rightarrow \infty$ limit. In the following, we
will mainly restrict ourselves to $\Theta =\Theta_{R}$ since we believe that
it is the correct solution. However all calculations have been redone for
$\Theta =\Theta_{eq}$ with related results.

\subsubsection{Numerical solution}
\begin{figure}[ht]
\[
\fig{6cm}{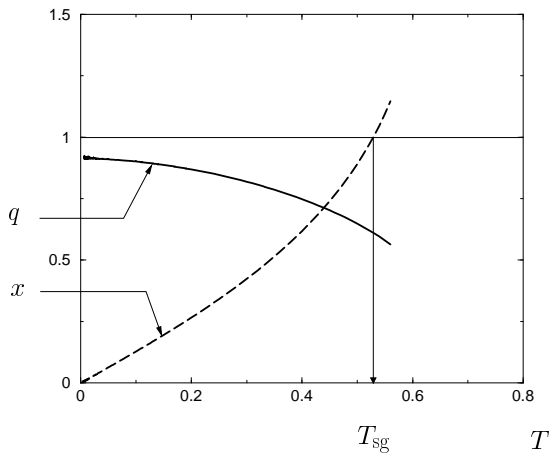}
\]
\caption{\label{TcFig} \sl The Edwards-Anderson parameter $q_{EA}$ and the
breakpoint $x$ as a function of the temperature $T$ for a fixed value
of the size of the spin $S=1$ ($J=1$). The transition to the
paramagnet is given by the condition $x=1$.}
\end{figure}

First, the critical temperature, is obtained from the numerical solution of
Eqs.(\ref{EqSG}) in imaginary time : the spin glass order
parameter $q (T) = g^{2} (T)$ and the breakpoint $x (T)$ are displayed
in Figure \ref{TcFig} as a function of the temperature.
$x$ increases linearly with $T$ from 0 at $T=0$ (there is no
replica symmetry breaking at zero temperature) and  $T_{sg}$ is determined
by the condition $x (T_{sg})=1$, since we must have $0\leq x\leq 1$ by
definition \cite{MezardParisiBook}.
Hence there is a discontinuity in $q$ at the transition, but we will
show below that the transition is second order.
A careful numerical study shows that the transition is always driven
by $x=1$ for all values of $S$ and produces  the critical temperature
displayed on Figure \ref{DiagrammePhase}. The computation is similar
for the critical temperature $T^{c}_{sg}$ given by the ``replicon''
criterion and for the critical temperature $T^{eq}_{sg}$ given  by
the equilibrium criterion.  Moreover,
we find that  $T^{c}_{sg}$, the {\sl dynamic} transition temperature, is {\bf
higher} than $T^{eq}_{sg}$, the {\sl static} transition temperature :
this is required by our physical interpretation of the two solutions
but it was  not obvious a priori from the integral equations solved.

\subsubsection{The large-$S$ limit and spectral densities}
\label{sec:largeS}

Further analytical insight into the spin glass phase itself can be obtained  by
considering various large-$S$ limits, which differ by the manner in which
temperatures and frequencies are scaled with $S$ (See
Fig~\ref{DiagrammePhase}).

In a first simple large-$S$ limit, we take $T$ large enough so
that $T/JS^2$ is of order unity. This is the simple classical
limit in which we can neglect all non-zero Matsubara frequencies,
and (\ref{DefSachdevYe}) reduces to the classical problem in which
$\vec{S}$ are commuting vectors of length $S$. The equations
(\ref{EqBase}) are analytically solvable and we can obtain a
closed form expression for the critical temperature at which
spin-glass order vanishes:
\begin{equation}\label{TsgClass}
T^{c}_{sg} \sim \frac{2}{3\sqrt{3}} JS^{2}.
\end{equation}

A second, more sophisticated limit, valid at lower temperatures (well within
the spin-glass phase) is when we examine $\omega$ and $T$ of order
$JS$. It is therefore useful to define the variables
 $\overline{\omega} =\omega /(JS)$ and $\overline{T}=T/(JS)$, which remain of order
 unity at large $S$.
 With this scaling,
the integral equations (\ref{EqSG}) reduce to independent quartic equations for
each frequency.
More precisely, if we make the following Ansatz for the Green's
function
\begin{equation}\label{DefScalingLimitLargeS}
\widetilde{G} (\omega ,T) = \frac{1}{JS} g_{1}  (\overline
{\omega} ,\overline{T}) + \frac{1}{JS^{2}} g_{2}
(\overline{\omega} ,\overline{T}) + \dots,
\end{equation}
then to leading order in $1/S$, (\ref{EqSG}) reduce to
\begin{eqnarray}\label{EqQuartic}
g (T) &=& S - \int_{-\infty }^{\infty} \rho_{1} (\overline{\omega}
) d\overline{\omega} \nonumber \\
g_{1} (\overline{\omega} )^{-1} &=& \overline{\omega}  -
\frac{1}{\Theta }- 3\Theta - 2g_{1} (\overline{\omega} ) -
\overline{g_{1} (-\overline{\omega})}
\end{eqnarray}
where $\rho_{1} =- \Im g_{1}/\pi $ as usual. Eliminating the
frequency $-\overline{\omega} $, we find a quartic equation for
$g_{1} (\overline{\omega} )$. We do not explicitly display the far
more complicated equation for the subleading term $g_2$.

\begin{figure}[ht]
\[
\fig{5cm}{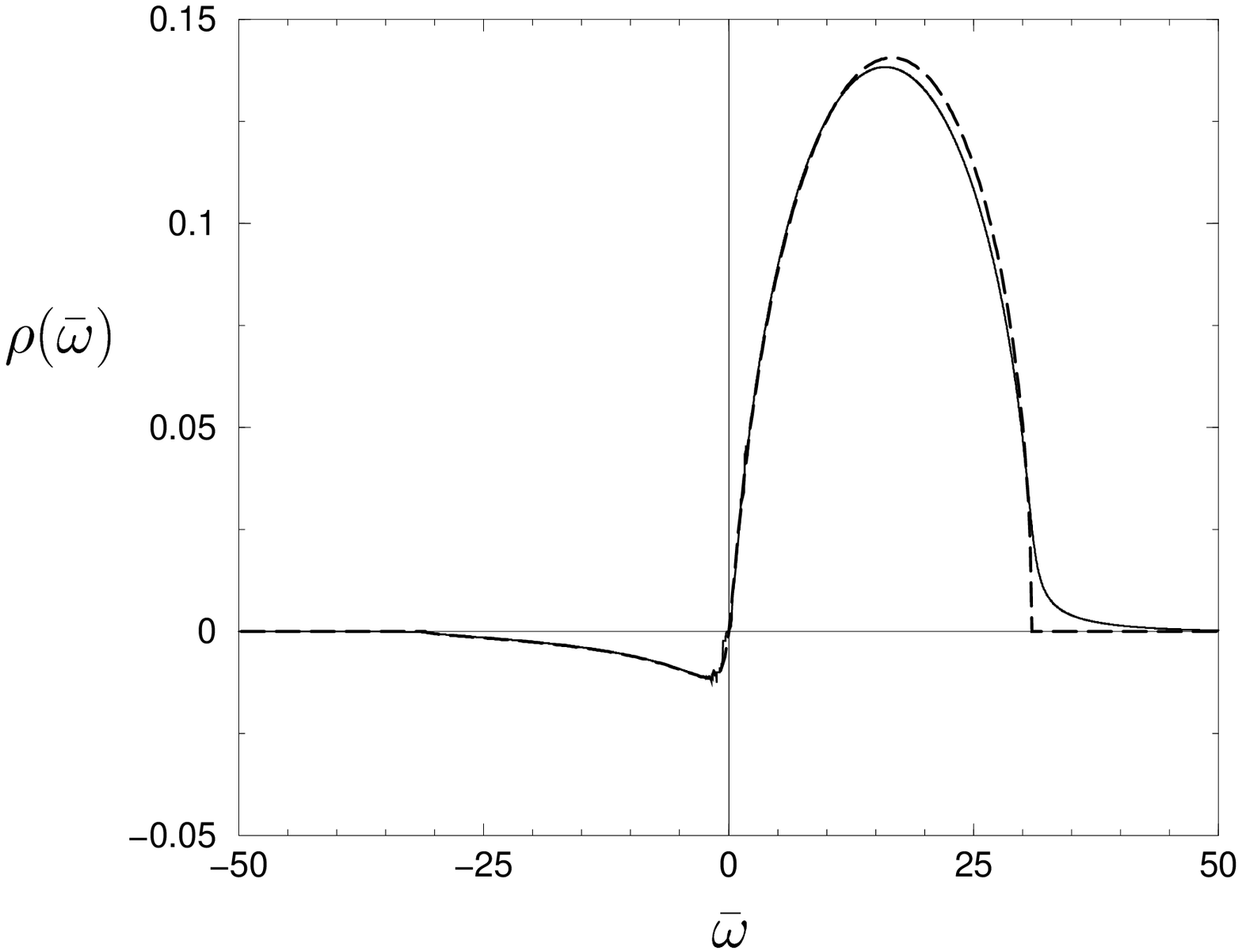} \kern  2cm \fig{6cm}{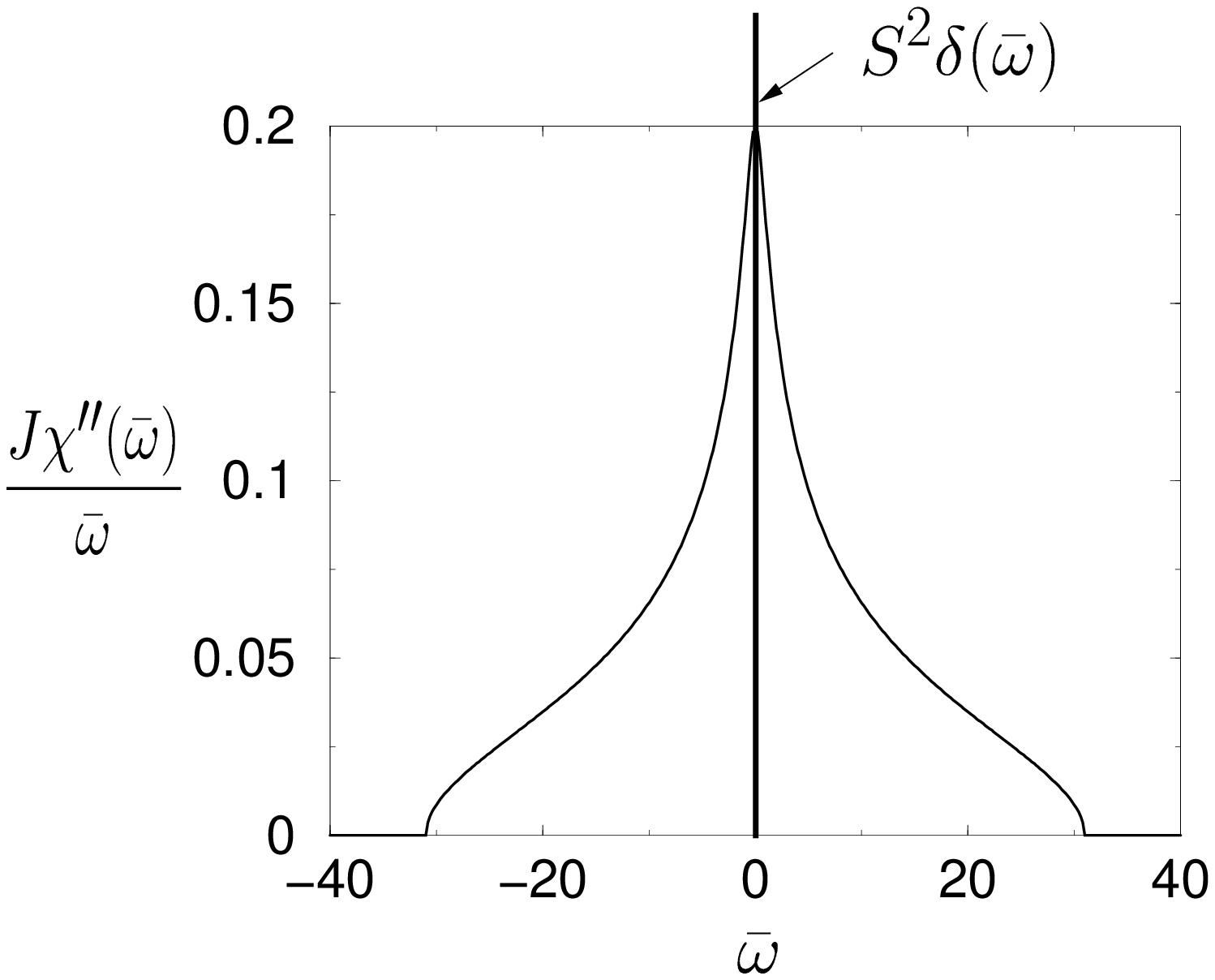}
\]
\caption{\label{FigRho} \sl
{\sl a)} $\rho_{1} (\omega )$ at $T\approx 0$ for $S=5$ and
$\Theta=\Theta_R$ :  the solid line
is the numerical
solution for the integral equation (\ref{EqSG}), the dashed line is
the solution of the quartic equation (\ref{DefScalingLimitLargeS}) for
$g_{1}$.
{\sl b)} $\chi'' (\omega )/\omega $ at $T\approx 0$ from the quartic
equation.}
\end{figure}

The solution of the quartic equation for $\Theta=\Theta_R$
is presented on Figure
\ref{FigRho}  together with a numerical solution of the full integral
equation for $S=5$.
From the solution of the quartic equation we find that $\rho_1$
vanishes linearly frequency at low frequencies; indeed, we find the analytic expansion
\begin{equation}
g_1 (\overline{\omega}) = - \frac{1}{\sqrt{3}} - \frac{(1+i)}{2}
\overline{\omega} + \frac{(2-3i)\sqrt{3}}{4} \overline{\omega}^2
\label{ss10}
\end{equation}
at low frequencies. We expect that the full Green's function in
(\ref{DefScalingLimitLargeS}) also has a similar low frequency
expansion, although 
this has not been proved. It is not difficult to show that the linear
low frequency 
spectral density holds at higher orders in the $1/S$ expansion;
moreover, our numerical 
results, shown in Fig~\ref{FigRho}, also clearly indicate a linear
behavior at small $\omega$.  At dominant order in the present large $S$ theory, the spin susceptibility is given by :
\begin{equation}\label{KiInSpinGlass}
\chi'' (\omega ) = - \pi \int_{0}^{\omega } \, dx \rho_{1} (x)
\rho_{1} (x-\omega )
+ g\pi \left(\rho_{1} (\omega ) - \rho_{1} (-\omega ) \right) + \pi
g^{2} \beta \omega 
\delta (\omega )
\end{equation}
and this is also shown in Figure \ref{FigRho}.

It is important to realize that the deceptively simple structure
in (\ref{ss10}) relies on the special value $\Theta=\Theta_R =
1/\sqrt{3}$ determined by the entirely different replicon
argument in Section~\ref{sec:replicon}. For arbitrary values of
$\Theta$ we either find no physically sensible solution of the
large $S$ quartic equation (this is the case at the replica
symmetric value $\Theta = 1$ where the spectral density does not
satisfy the required positivity criteria) or a solution with a
spectral gap. In the latter case, the solution for $g_1$ is real
for small real $\overline{\omega}$, and there is an onset in the
imaginary part $\sim (\overline{\omega} -
\overline{\omega}_c)^{1/2}$ above some critical frequency
$\overline{\omega}_c$. The solution for $\Theta=\Theta_{eq}$ is of
the second type: it has a finite energy gap, but does not violate
any spectral positivity criteria.

This subsection has so far identified two distinct large $S$ regimes.
In the regime $T \sim JS^2$ we have purely classical behavior (the non-zero
Matsubara frequencies can be neglected for static properties) and
a phase transition at a critical temperature in (\ref{TsgClass}) where
the spin-glass order vanishes. At lower temperatures, $T \sim JS$, we are well
within the spin-glass phase, and the semiclassical dynamics is described
by the solution of a quartic equation defined by (\ref{EqQuartic}).
As we noted in Fig~\ref{DiagrammePhase}, there is a third ``quantum''
regime at even lower temperatures, $T \sim J \sqrt{S}$, and this
becomes evident in a study of the thermodynamic properties presented
in the following section.

\subsection{Thermodynamics}\label{Thermo}

We now turn to the internal energy $U$ and the specific heat $C$.
By computing the average of the Hamiltonian in the $N\rightarrow
\infty $ limit, we find that $U$ is given by :
\begin{equation}\label{FormulaU}
U (T) = -\frac{J^{2}}{2}\int_{0}^{\beta } d \tau \left[
G^{ab}(\tau ) G^{ab} (-\tau ) \right]^2
\end{equation}
Using (\ref{AnsatzParisi}) and the one-step replica symmetry
breaking Ansatz in the spin glass phase, we find :
\begin{equation}\label{FormuleDeveloppeeU}
U (T)  = \left\{
\begin{aligned}
-&\frac{J^{2}}{2}\int_{0}^{\beta } G(\tau )^{2}G(-\tau )^{2}d\tau
& \qquad \text{ in the paramagnetic phase} \\
 -&\frac{J^{2}}{2}\int_{0}^{\beta }
\bigl (\widetilde{G}(\tau) -g \bigr )^{2} \bigl
(\widetilde{G}(-\tau )-g \bigr )^{2}d\tau - \frac{J^{2}}{2} \beta
(x-1)g^{4}
&   \qquad \text{ in the spin glass phase} \\
\end{aligned}
\right.
\end{equation}

\begin{figure}[ht]
\[
\fig{8cm}{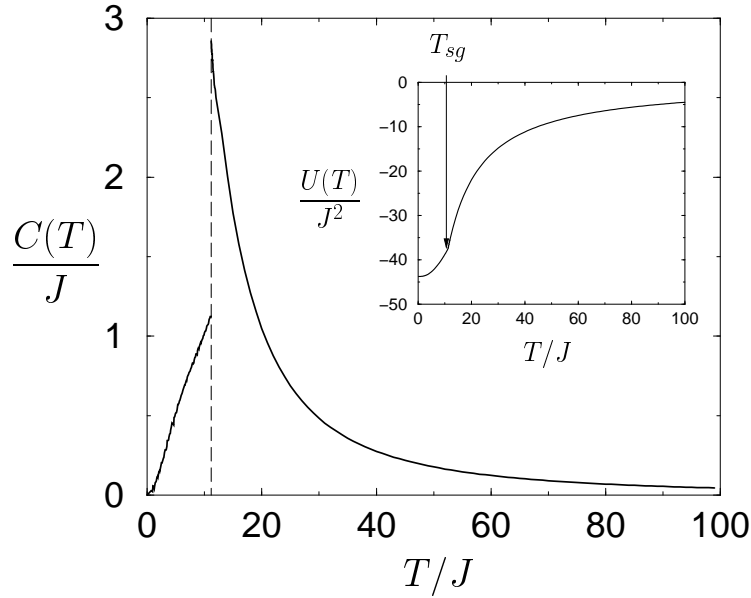}
\]
\caption{\label{CetU} \sl The specific heat $C (T)$ and the
internal energy $U (T)$ vs. the temperature $T$, from a numerical
solution of Eqs. (\ref{EqSG}) for $S=5$ and $\Theta=\Theta_R$. }
\end{figure}

A numerical computation of the internal energy $U (T)$ and the
specific heat $C (T)$ is displayed in Figure \ref{CetU} for
$\Theta=\Theta_R$.
The condition for the phase transition between the two phases is
that the breakpoint in the Parisi function reaches its limiting
value $x=1$. In this limit the equations determining the
parameters in the spin glass phase, (\ref{EqSG}), transform
continuously to those for the paramagnet. As the equations are
believed to have a unique solution, this implies that there is no
discontinuity in the internal energy at the transition, indicating
its second-order nature. This is confirmed by the numerical
solution displayed on Figure \ref{CetU}.

Moreover, in the large-$S$ limit defined above, we can perform a
low temperature expansion of the internal energy. Inserting
(\ref{DefScalingLimitLargeS}) and (\ref{EqQuartic}) into
(\ref{FormulaU}), and keeping the dominant terms at large $S$ we
find
\begin{eqnarray}
U(T) &=& - \frac{J}{2} \left( \frac{1}{\Theta} + 3 \Theta \right)
g^2 -J^2 g^2 \frac{1}{\beta} \sum_{\nu_n} \left[ \widetilde{G} (
i \nu_n ) \widetilde{G} ( - i \nu_n) + 2 \widetilde{G}^2 (i \nu_n
)
\right] + \ldots \nonumber \\
&=& - \frac{JS^2 }{2} \left( \frac{1}{\Theta} + 3 \Theta \right)
+ J S \left( \frac{1}{\Theta} + 3 \Theta \right)  \overline{T}
\sum_{\overline{\nu}_n} g_1 ( i \overline{\nu}_n) -J S
\overline{T} \sum_{\overline{\nu}_n} \left[ g_1 (i
\overline{\nu}_n ) g_1 ( - i \overline{\nu}_n) + 2 g_1^2 ( i
\overline{\nu}_n ) \right] + {\cal O} (JS^0) \label{ssU}
\end{eqnarray}
where $\overline{\nu}_n = \nu_n / (JS)$. Clearly, this result
indicates that the leading term in $U(T)$ is a
temperature-independent constant of order $JS^2$, followed by a
term of order $JS$ whose coefficient is a function only of
$(T/JS)$. Evaluation of the latter function at low $T$ for
$\Theta=\Theta_R$ yields a curious accident: the gapless
structure of the spectral functions suggests that the low $T$
expansion should depend only on even powers of $T/JS$, but it is
not difficult to show using (\ref{ss10}) that the coefficient of
the term of order $S(T/JS)^2$ vanishes. The first non-vanishing,
$T$-dependent term among those shown explicitly in (\ref{ssU})
turns out to be order $JS (T/JS)^4$. To obtain the true low $T$
behavior we need to expand (\ref{ssU}) to one higher-order in
$1/S$, and this requires use of the second term, $g_2$, in
(\ref{DefScalingLimitLargeS}). We do not expect any cancellation
of the term of order $(T/JS)^2$ at this point, and so the low $T$
expansion for $U$ looks like
\begin{equation}\label{DevEnergy}
U(T) = U(0) + a S (T/JS)^{4} + b (T/JS)^{2} +\dots
\end{equation}
Rather than numerically evaluating the values $a$ and $b$, we
will be satisfied by the full numerical solution of (\ref{EqSG}),
followed by the evaluation of (\ref{FormuleDeveloppeeU}). The
results are shown in Fig~\ref{CetU} and are consistent with
(\ref{DevEnergy}). The structure of the expansion in
(\ref{DevEnergy}) suggests that these results are valid for $T < J
\sqrt{S}$, where the specific heat depends linearly on the
temperature. Although the present discussion has been carried out
for large $S$, we expect, and this is supported by our numerical
results, that the linear $T$ dependence of the specific heat holds
even for small $S$ as $T \rightarrow 0$.

In Appendix~\ref{rotor} we describe the computation of the
specific heat of the quantum rotor and Ising spin glasses
considered in Ref.~\onlinecite{rsy}. As noted in the introduction,
these models are simpler because they do not have quantum Berry
phases in their effective action. Further, at low orders in their
Landau theory, the solution for the spin glass phase is
replica-symmetric. However, understanding the true $T \rightarrow
0$ behavior requires inclusion of higher-order, ``dangerously
irrelevant'' terms which induce replica symmetry breaking; this
is carried out in Appendix~\ref{rotor}, and we find that these
quantum spin glasses also have a linear specific heat at low $T$.

\section{Conclusion}\label{conclusion}
We believe that the results of this paper provide a reasonably
complete understanding of the infinite-range quantum Heisenberg
spin glass. While there have been a large number of previous
studies of quantum spin glasses of Ising spins and rotors
(including models with $(p>2)$-spin interactions), none of these
models contain quantum Berry phases in their effective actions,
as is the case with the Heisenberg model.
They have strong consequences: the spin-liquid solution of
Section~\ref{SubSectSpinliquid} and its spectral density
(\ref{FormeEchelle}) are novel properties of the Heisenberg
model. There is an intricate interplay in stability between this
spin-liquid state and the state with spin-glass order at low $T$
which we have also described. At sufficiently low $T$, the
spin-glass order always appears, and we have also described the
thermodynamic properties of this state.

An important issue not resolved in our analysis is the origin of
the marginal stability criterion in the fluctuation eigenvalues
in replica space. We imposed this criterion in a rather {\em ad
hoc} manner, and found that it was the unique case under which the
quantum excitation spectrum was gapless. Ultimately, the
selection criterion for the spin glass state has to be a dynamic
one, and this requires an analysis of the approach to equilibrium
in real-time dynamics. Such an analysis was not carried out here,
and is an important direction for future research.

Another interesting open problem is to extend the study of
(\ref{DefSachdevYe}) to cases where $J_{ij}$ has a non-zero
average value. This will allow for ground states with other types
of magnetic order, ferromagnetic and antiferromagnetic, and their
competition with the spin glass state should be of some
experimental interest. Interesting transitions in the
paramagnetic states from the spin liquid state discussed also
appear possible.

We have already mentioned a recent study \cite{Slush} of the
quenching of the spin liquid state by mobile charge carriers into
a disordered Fermi liquid. Combining this with models just
mentioned, with a non-zero average $J_{ij}$, should lead to
results of direct physical interest in the heavy fermion and
cuprate series of compounds.

\begin{acknowledgements}
We thank G. Biroli, L. Cugliandolo, D. Grempel, P. Le Doussal,
M. Rozenberg for useful discussions.
S.S. was supported by US NSF Grant No DMR 96--23181.
O.P is supported by the Center of Material Theory, Rutgers University, NJ,USA.
\end{acknowledgements}

\begingroup
\appendix

\section{Computation of the spectral asymmetry}\label{AppSpectralAsym}
This appendix is devoted to the derivation of Eqs
(\ref{ValueOfTheta}).
 We will consider
hereafter the fermionic case (the bosonic one is very similar).
At zero temperature, the number of particles is given by
\begin{equation}\label{NbrePart}
q_{0} =i \int_{-\infty }^{\infty } \frac{d\omega}{2\pi }  \,\,
G_{f}^{F}(\omega )e^{i \omega 0^{+}}
=i \ppint_{-\infty }^{\infty } \frac{d\omega}{2\pi }  \,\,
\partial_{\omega } \ln  G_{f}^{F} (\omega ) e^{i \omega 0^{+}}
- i \ppint_{-\infty }^{\infty } \frac{d\omega}{2\pi }  \,\,
G_{f}^{F} (\omega ) \partial_{\omega }\Sigma_{f}^{F} (\omega)
 e^{i \omega 0^{+}}
\end{equation}
where $G^{F}$ is the Green function with Feynman prescription on the
real axis, and the symmetric principal part is defined by
\begin{equation}\label{DefPP}
\ppint_{-\infty}^{\infty} = \lim_{\eta \rightarrow 0 }{
\int_{-\infty}^{-\eta} + \int_{\eta }^{\infty }}
\end{equation}
Using the relation between $G^{F}$ and the {\it retarded Green
function} $G^{R}$, we find for the first term :
\begin{equation}\label{Terme1}
i \ppint_{-\infty }^{\infty } \frac{d\omega}{2\pi }  \,\,
\partial_{\omega } \ln  G_{f}^{F} (\omega ) e^{i \omega 0^{+}}
= \frac{\arg G_{f}^{R} (0^{-}) - \arg G_{f}^{R} (-\infty)}{\pi} +
i \ppint_{-\infty }^{\infty } \frac{d\omega}{2\pi }  \,\,
\partial_{\omega } \ln  G_{f}^{R} (\omega ) e^{i \omega 0^{+}}
\end{equation}
The arguments can be extracted from the low-energy and the high-energy
behavior of the Green function, which leads to $\arg G_{f}^{R}
(0^{-}) = -3 \pi /4 -\theta$ and $\arg G_{f}^{R} (-\infty) = -\pi $
respectively. The  integral on the right of (\ref{Terme1}) can be
easily evaluated : we close the contour of integration, avoiding  the
singularity at $\omega =0$ and use the analyticity of the retarded
Green function in the upper half plane, in which it has no zeros nor
poles.
We find finally :
\begin{equation}\label{Finalq01}
q_{0} = \frac{1}{2} - \frac{\theta }{\pi} -
 i \ppint_{-\infty }^{\infty } \frac{d\omega}{2\pi }  \,\,
G_{f}^{F} (\omega ) \partial_{\omega }\Sigma_{f}^{F} (\omega)
 e^{i \omega 0^{+}}
\end{equation}

The problem is now reduced to the computation of the integral in
(\ref{Finalq01}) as function of $\theta$, which turn out to be
the most difficult point. An analogous computation was performed
in the overscreened regime of a large-$N$ description of Kondo
effects \cite{KondoPRL,KondoLong}, but it turns out to be more
complex here. Proceeding along the lines of
Ref.~\onlinecite{KondoPRL,KondoLong}, we note the existence of
Luttinger-Ward functional
 $\Phi_{LW} = \int dt \,G^{2} (t)G^{2} (-t)$
which has two properties : first we have $\Sigma^{R} (\omega ) =
\delta \Phi_{LW} /\delta G^{R} (\omega )$; second $\Phi_{LW }$ is invariant
 in the  transformation $G (\omega )\rightarrow G (\omega +\epsilon)$.
From this, we could naively think that the integral of
(\ref{Finalq01}) vanishes. However, it is not possible to find a
regularization for the integral for which we could use the invariance
of the Luttinger-Ward functional and a more careful analysis shows that :
\begin{equation}\label{Anomalie}
i \ppint_{-\infty }^{\infty } \frac{d\omega}{2\pi }  \,\,
G_{f}^{F} (\omega ) \partial_{\omega }\Sigma_{f}^{F} (\omega)
= \frac{\sin 2\theta }{4}
\end{equation}

To obtain this result, we introduce the following parametrisation of
the singularity at $\omega =0$ :
\begin{equation}\label{DefC+C-}
\rho (\omega ) \sim
\left\{
\begin{aligned}
\frac{C_{+}}{\sqrt{\omega}}  &\text{\qquad \qquad for $\omega >0$}\\
\frac{C_{-}}{\sqrt{|\omega|}} &\text{\qquad \qquad for $\omega <0$}
\end{aligned}
\right.
\end{equation}
The principle of the computation is very simple : we  compute
explicitly the integral  with a regulator $\eta>0$ and
then perform the limit $\eta \rightarrow 0$.
Going to the real axis, we find :
\begin{equation}\label{ExprSigF}
\Sigma^{F}(\omega ) = -
\int_{\substack{\\ \\
\omega_{1}>0\\
\omega_{2}>0\\
\omega_{3}<0\\
}
\substack{\\ \\ \\ \text{ ou }\\ \\}
\substack{\\ \\
\omega_{1}<0\\
\omega_{2}<0\\
\omega_{3}>0\\
}
}
d \omega_{1} d \omega_{2} d \omega_{3} \,\,
\frac{\rho (\omega_{1})\rho (\omega_{2})\rho (\omega_{3})}{
\omega_{1} + \omega_{2}-\omega_{3} -\omega -i0^{+}\sgn \omega_{1}}
\end{equation}
Using the notations  $a = \overline{a} +  i\epsilon_{a}$, $b =
\overline{b} + i\epsilon_{b}$, $\epsilon_{a/b} = \pm 0^{+}$ and
$\psi_{\eta }(x) = \Theta \left(|x| - \eta  \right)$
($\overline{a}$ are $\overline{b} $ real and $\Theta$ is the
Heaviside function), we obtain (using the definition of the
principal part (\ref{DefPP}))
\begin{align}\label{Utile1}
\phi_{\eta } (a,b) = &  \ppint_{-\infty}^{\infty} \frac{dz}{\left(z-a
\right)^{2} (z-b)}\\ 
\nonumber
 = & \frac{1}{(a-b)^{2}}
\left(
\ln \left| \frac{(\eta +b) (\eta -a)}{(\eta -b) (\eta +a)} \right|
+ i\pi \psi_{\eta } (b) \sgn \epsilon_{b}  - i\pi  \psi_{\eta }(a)
\sgn \epsilon_{a}
\right)
  + \frac{1}{a-b}\left(\frac{1}{\eta -a} + \frac{1}{\eta +a} \right)
\end{align}
Using the spectral representation for  $G^{F}$ and (\ref{ExprSigF}),
we find :
\begin{equation}\label{R1}
{\cal  I} = -
\int_{\Delta_{1}\cup \Delta_{2}}^{}\prod_{k=0}^{3} d \omega_{k} \,\,
\rho (\omega_{0})\rho (\omega_{1})\rho (\omega_{2})\rho (\omega_{3})
\phi_{\eta} (\omega_{1} +\omega_{2}-\omega_{3}-i\epsilon_{1}\sgn
\omega_{1}, \omega_{0} - i\epsilon_{0}\sgn \omega_{0})
\end{equation}
with an explicit integration over  $\omega $ with  (\ref{Utile1}).
In this  expression, the integration domains are defined as :
\begin{equation}\label{Domaines}
\Delta_{1} =
\left\{\begin{aligned}
\omega_{0}&<0\\
\omega_{1}&>0\\
\omega_{2}&>0\\
\omega_{3}&<0
\end{aligned}\right\}
\bigcup
\left\{\begin{aligned}
\omega_{0}&>0\\
\omega_{1}&<0\\
\omega_{2}&<0\\
\omega_{3}&>0
\end{aligned}\right\}
\qquad \qquad
\Delta_{2} =
\left\{\begin{aligned}
\omega_{0}&>0\\
\omega_{1}&>0\\
\omega_{2}&>0\\
\omega_{3}&<0
\end{aligned}\right\}
\bigcup
\left\{\begin{aligned}
\omega_{0}&<0\\
\omega_{1}&<0\\
\omega_{2}&<0\\
\omega_{3}&>0
\end{aligned}\right\}
\end{equation}
Since  $\phi_{\eta } (-a,-b) = - \phi_{\eta } (a,b)$,
a simple change of  variable leads to :
\begin{align}\label{R2}
\nonumber
{\cal I} &=
-\int_{x_{i}>0}
\Biggl[
\Bigl(
\rho(\omega_{1})\rho(\omega_{2})\check\rho(\omega_{3})\check\rho(\omega_{0})-
\check\rho(\omega_{1})\check\rho(\omega_{2})\rho(\omega_{3})\rho(\omega_{0})
\Bigr)
\phi_{\eta} (x_{1}+x_{2}+x_{3} - i\epsilon_{1},-x_{0}+i\epsilon_{0})\\
& + \Bigl(
\rho(\omega_{1})\rho(\omega_{2})\check\rho(\omega_{3})\rho(\omega_{0})-
\check\rho(\omega_{1})\check\rho(\omega_{2})\rho(\omega_{3})
\check\rho(\omega_{0})   
\Bigr)
\phi_{\eta} (x_{1}+x_{2}+x_{3} - i\epsilon_{1},x_{0}-i\epsilon_{0})
\Biggr]
\end{align}
with $\check \rho (\omega ) = \rho (-\omega )$.
To take the limit  $\eta \rightarrow 0$, we use the new variables
$x_{i}=\eta u_{i}$ and the behavior of  $\rho (x)$ for $x\rightarrow
0$, parametrised according to (\ref{DefC+C-}).
The first integral in  (\ref{R2}) vanishes at dominant order in
$\eta$ (this term is proportional to   $C_{+}^{2}C_{-}^{2}
-C_{-}^{2}C_{+}^{2}=0 $), but the second integral gives :
\begin{equation}\label{R3}
{\cal I}= \int_{u_{i}>0} \frac{C_{+}^{3}C_{-} -
C_{-}^{3}C_{+}}{\sqrt{u_{0}u_{1}u_{2}u_{3}}}
\phi_{\eta=1} (u_{1}+ u_{2} + u_{3} -i\epsilon_{1},u_{0}-i\epsilon_{0})
\end{equation}
Using  $x=u_{0}$, $y=u_{1}+u_{2}+u_{3}$ and polar coordinates in
$\sqrt{u_{i}}$, we find
 ${\cal I} = 2\pi (C_{+}^{3}C_{-} - C_{-}^{3}C_{+}) {\cal I}_{2 }$
with
\begin{multline}\label{R4}
{\cal I}_{2} =  \int_{0}^{\infty }\!\!\!\! \int_{0}^{\infty }
\frac{dx}{\sqrt{x}} \sqrt{y}dy
\left[ \frac{1}{(x-y+i\epsilon )^{2}}
 \left(
      \ln \left| \frac{1+x}{1-x}\frac{1-y}{1+y} \right|
        +  i \pi \left(\psi_{1} (y) - \psi_{1} (x) \right)
 \right) +   \right.\\
\left. \frac{1}{y-x-i\epsilon }
  \left(\frac{1}{1-y+i\epsilon_{1}} + \frac{1}{1+y-i\epsilon_{1}} \right)
\right]
\end{multline}
After an integration by parts on $y$ and using 
\begin{equation}\label{Utile2}
\int_{0}^{1} \frac{dx}{x} \ln \left|\frac{1+x}{1-x} \right| = \frac{\pi^{2}}{4}
\end{equation}
we find
\begin{equation}\label{R7}
i\ppint_{-\infty }^{\infty } \frac{d\omega}{2\pi}  \,\,
G_{f}^{F} (\omega ) \partial_{\omega }\Sigma_{f}^{F} (\omega)  =
\frac{\pi^{3}}{2}(C_{+}^{3}C_{-} - C_{-}^{3}C_{+})
\end{equation}
Finally, a analogous computation can be performed in the bosonic case,
leading in both cases to :
\begin{equation}\label{FinalThetaPre}
i\ppint_{-\infty }^{\infty} \frac{d\omega}{2\pi}  \,\,
G^{F} (\omega ) \partial_{\omega }\Sigma^{F} (\omega)  =
\frac{\sin 2\theta }{4}
\end{equation}
(in this expression, $-\pi \leq\theta \leq \pi $).
These expressions have been shown to agree perfectly with  numerical
computations in
imaginary time for the fermionic case and on the real axis at zero temperature
in the bosonic case.

Let us note finally that we can guess the result if we admit a priori
that the integral is given
by an homogeneous polynomial of degree 4 : due to the particle-hole
symmetry (in the fermionic case : $f\leftrightarrow f^{\dagger}$, the
result can be expressed as a function of $C^{4}_{+}-C^{4}_{-}$ and
$C_{+}^{3}C_{-} - C_{-}^{3}C_{+}$. The first term is rejected since it
leads to a singularity at $\theta =\pm \pi /4$. The proportionality
coefficient is fixed by imposing $\theta =\pi /4$ for $q_{0}=0$.

\section{The marginality criterion}\label{AppReplicon}
\subsection{Diagonalization of the fluctuation matrix}
First, we diagonalize the fluctuation matrix $M$ in the
replica space defined by Eq. (\ref{DefMatrixFluctuations}).
A priori $M$ is a $n (n-1)/2\times n (n-1)/2$ matrix. However, we have
taken for $g_{ab}$ the simple one step replica symmetry breaking
Ansatz on $g_{ab}$, {\it i.e.} the $n\times n $ matrix splits into
$n/m\times n/m$ blocks : $g_{ab}=g$ if $\lfloor a/m \rfloor  = \lfloor
b/m \rfloor $, 0 otherwise.
Thus $M$ splits into $n/m$ identical $m (m-1)/2\times m (m-1)/2$
blocks ($M_{ab,cd}$ does not vanish if and only if all indices are in
the same block
$\lfloor a/m \rfloor  = \lfloor b/m \rfloor =\lfloor c/m \rfloor  = \lfloor
d/m \rfloor $).
Hence the diagonalization is to be performed only on one block (we set
$1\leq a,b,c,d \leq m$), which elements are given by (with $a,b,c,d$
distinct replica indices) :
\begin{align}\label{AppRepValueM}
M_{ab,ab}&=A \equiv 3 \beta J^{2}g^{2} \left[1-3
\beta^{2}J^{2}g^{2}\left (g^{2} +\Bigl(g + \frac{\Theta }{\beta Jg}
\Bigr)^{2}\right) \right] \\
M_{ab,ac}&=B\equiv
-9\beta ^{3}J^{4}g^{5} \left(2g +\frac{\Theta }{\beta J g} \right)\\
M_{ab,cd}&=C\equiv  -18\beta^{3}J^{4}g^{6}
\end{align}
This matrix has already been diagonalized in
Ref.~\onlinecite{DeAlmeida} and its eigenvalues are given by :
\begin{align}\label{AppRepValeurde_e}
e_{1} &= A-2B+C \\
e_{2} &= A +2 (m-2)B + (m-2) (m-3)C/2\\
e_{3} &= A + (m-4) B  - (m-3)C
\end{align}
Moreover, the degeneracies of $e_{1}$, $e_{2}$ and $e_{3}$ are
$n (m-3)/2$, $n/m$ and $n (m-1)/m$ respectively.
Using (\ref{AppRepValueM}) and (\ref{AppRepValeurde_e}), we find
finally the result of the text (\ref{EigenvaluesFluct}).

The above calculation has entirely ignored perturbations,
$\delta \widetilde{G}(\tau)$
in the
diagonal elements of (\ref{AnsatzParisi}). Including these greatly
complicates the analysis, but a simple observation will suffice
for our purposes. Our main attention is on the cross-coupling
between $\delta \widetilde{G}(\tau)$ and the $\delta g_{ab}$. A
simple consequence of the block-diagonal structure of the $g_{ab}$
in the mean-field solution is that this cross-coupling has the
form
\begin{equation}
\delta {\cal F} \sim \sum_{a>b, \lfloor a/m \rfloor  = \lfloor
b/m \rfloor} \int_0^{\beta} d \tau \delta \widetilde{G}
(\tau) \delta g_{ab}.
\label{ssapp}
\end{equation}
Now we can expand $\delta g_{ab}$ in terms of the eigenvectors
associated with (\ref{AppRepValeurde_e}), which were computed in
Ref.~\onlinecite{DeAlmeida}. The key observation is that after
the sum over $a,b$ in (\ref{ssapp}), the cross-terms
corresponding to all the eigenvectors associated with $e_{1}$
vanish. Consequently these eigenvectors remain eigenvectors even
upon including $\delta \widetilde{G} (\tau)$, and the eigenvalue
$e_1$ remains unchanged. A similar argument shows that the
eigenvalue $e_3$ also remains unchanged, and only the
eigenvectors associated with $e_2$ are modified non-trivially by
the coupling to $\delta \widetilde{G} (\tau)$.

\subsection{The replicon solution is gapless}
\label{app:gapless}

Let us assume that there is no gap in the boson spectral density and
more precisely that for small $\omega$ :
\begin{equation}\label{AppGaplessAnsatz}
\widetilde{G} (\omega ) = \frac{\Theta }{Jg} + (a+ib)\omega^{\alpha}
\Theta (\omega ) + (a'+ib')|\omega|^{\alpha}\Theta (-\omega ) + o
(|\omega|^{\alpha})
\end{equation}
where $\Theta $ is the Heaviside function, $a,b,a',b'$ are real
constants, and $\alpha >0$.
Then from (\ref{EqSG_DefSigTilde}) we obtain for $\omega >0$ :
\begin{align}
\Im  \left(\widetilde{\Sigma } (\omega )\right)  &= J^{2}g^{2}
\Im \left( 2 \widetilde{G} (\omega ) + \overline{\widetilde{G}
(-\omega )} \right) + \dots \\
\widetilde{\Sigma } (\omega ) &= c+ \left(d+ i (2b-b')J^{2}g^{2}
\right) \omega^{\alpha}
\end{align}
where $c$ and $d$ are real constants.
The other terms are subdominant in the limit
$\omega \rightarrow 0$ as can be seen using a spectral representation.
We then expand (\ref{EqSG_Dyson}) to second order and obtain
at first order $\lambda =c +\frac{Jg}{\Theta }$ and for the imaginary
part at second order :
\begin{equation}
b - (2b-b')\Theta^{2} = b' - (2b'-b)\Theta^{2}=0
\end{equation}
which leads to ($\Theta =1$ is excluded since $b>0$ and $b'<0$)
\begin{equation}
\Theta^{2}=\frac{1}{3}
\end{equation}
Thus the value of $\Theta $ given by the replicon
condition is {\sl the only one} that leads to a gapless bosonic
spectral density.

\section{Free energy of quantum rotor and Ising spin glasses}
\label{rotor}

Quantum spin glasses of quantum rotors and Ising spins were
studied extensively in Ref.~\onlinecite{rsy}. However, while the
paramagnetic phase and the vicinity of the quantum-critical point
were fairly completely described, the $T \rightarrow 0$
thermodynamics within the spin glass phases were only studied in
the replica-symmetric solution. A proper understanding of this
low $T$ limit requires consideration of replica symmetry
breaking, and we will provide that here. We will find, as in the
more complex Heisenberg spin model considered in the body of the
paper, that the specific heat is linear in $T$ at low $T$.

As we are restricting our attention to mean-field theory, we can
neglect the spatial dependence of all degrees of freedom.
Further, we will also restrict ourselves to the Ising case, and
the generalization to the multi-component rotor case is
immediate. As discussed in Ref.~\onlinecite{rsy}, the effective
action of the quantum Ising spin glass is expressed in terms of
the order parameter functional
\begin{equation}
Q^{ab} (\tau, \tau^{\prime} )  = \langle \sigma^a (\tau) \sigma^b
(\tau^{\prime} ) \rangle \label{rsy1}
\end{equation}
where $\sigma^a$ is the Ising spin in replica $a$. The important
low-order terms in the free energy density are
\begin{eqnarray}
{\mathcal F} &=& \frac{1}{\kappa} \int d \tau \sum_a \left. \left[
\frac{\partial}{\partial \tau_1} \frac{\partial}{\partial \tau_2}
+ r \right] Q^{aa} (\tau_1, \tau_2 ) \right|_{\tau_1 = \tau_2 =
\tau} -\frac{\kappa}{3} \int d \tau_1 d \tau_2 d \tau_3 \sum_{abc}
Q^{ab} (\tau_1 , \tau_2 ) Q^{bc} (\tau_2, \tau_3) Q^{ca} (\tau_3,
\tau_1 ) \nonumber \\
&~&~~~~~~~+ \frac{u}{2} \int d \tau \sum_a Q^{aa} (\tau,\tau)
Q^{aa} (\tau,\tau) - \frac{y}{6} \int d \tau_1 d \tau_2 \sum_{ab}
\left[ Q^{ab} (\tau_1 , \tau_2 ) \right]^4. \label{rsy2}
\end{eqnarray}
Here $r$ is the parameter which tunes across the spin glass
transition, and $\kappa$, $y$ measure the strength of various
non-linearities. The analysis of thermodynamic properties in the
spin-glass phase  in Ref.~\onlinecite{rsy} was carried out with a
vanishing coefficient of the quartic term, $y=0$: in this case
the order parameter has replica symmetry, and it was found that
the specific heat $\sim T^3$ as $T \rightarrow 0$. Here, we will
extend the solution to small $y \neq 0$, and show that the
solution with broken replica symmetry has a linear specific heat.

Time-translational symmetry requires that the mean-field solution
take the form
\begin{equation}
Q^{ab} (\tau_1 , \tau_2 ) = \frac{1}{\beta} \sum_{\nu_n} Q^{ab}
(i \nu_n) e^{i \nu_n (\tau_1 - \tau_2 )}. \label{rsy3}
\end{equation}
As in (\ref{AnsatzParisi}), we choose the following Ansatz for
$Q^{ab}$:
\begin{equation}
Q^{ab} (i \nu_n) = \left\{ \begin{array}{ccc} D(i \nu_n) + \beta
q_{EA}
& \qquad & a=b \\
\beta q_{ab} & \qquad & a \neq b
\end{array}
\right. \label{rsy4}
\end{equation}
where the off-diagonal terms, $q_{ab}$, are time-independent and
characterized by the Parisi function $q(u)$, and $q(1) \equiv
q_{EA}$. We have included an additive factor of $\beta q_{EA}$ in
the diagonal term for convenience, and without loss of
generality: as in the discussion below (\ref{AnsatzParisi}), we
will find that this ensures that at $T=0$ the solution for
$D(\tau)$ vanishes as $\tau \rightarrow \infty$. Also, the
diagonal components $q_{aa}$ do not appear in the above, and we
are therefore free to choose them as $q^{aa} = 0$. Here, and in
the remainder of this appendix we are assuming that $r$ is
sufficiently negative so that the system has a spin glass ground
state; for larger $r$, the ground state is a paramagnet \cite{rsy}
with $q_{EA}=q_{ab} =0$ whose properties are not addressed here.

We now need to insert (\ref{rsy4}) into (\ref{rsy2}) and find the
saddle-point with respect to variations in the functions $q(u)$
and $D ( i \nu_n)$. This is, in principle, a straightforward
exercise, but the computations are somewhat lengthy.

We first identify just the terms that depend upon $q^{ab}$; these
have the form
\begin{equation}
{\mathcal F} =  - R_1 \mbox{Tr} q^2 - \frac{R_2}{3} \mbox{Tr} q^3
- \frac{R_3}{6} \sum_{ab} q_{ab}^4 + \ldots, \label{rsy5}
\end{equation}
where
\begin{eqnarray}
R_1 &=& \beta \kappa (D(0) + \beta q_{EA}) \nonumber \\
R_2 &=& \kappa \beta^2 \nonumber \\
R_3 &=& \beta y. \label{rsy6}
\end{eqnarray}
We first address the problem of determining the saddle-point of
(\ref{rsy5}) with respect to variations in $q(u)$. Fortunately,
this problem has been completely solved in the classical spin
glass literature \cite{HertzBook}, and we can directly borrow the
results: the function $q(u)$ increases linearly as a function of
$u$ for $0 < u < 1 - (1- 4 R_1 R_3/R_2^2)^{1/2}$, where it
saturates until $u=1$ at the constant value
\begin{equation}
q_{EA} = \frac{R_2 - (R_2^2 - 4 R_1 R_3)^{1/2}}{2 R_3}.
\label{rsy7}
\end{equation}
Combining (\ref{rsy7}) with (\ref{rsy5}), we obtain the simple
result
\begin{equation}
q_{EA}^2 = - \kappa D(0)/y \label{rsy8}
\end{equation}

Next, we consider the variation of ${\mathcal F}$ in (\ref{rsy2})
with respect to $D(i \nu_n)$. This is most easily done for $\nu_n
\neq 0$, for which we obtain the following saddle-point equation
\begin{eqnarray}
&& \frac{1}{\kappa} ( \nu_n^2 + r) - \kappa D^2 (i \nu_n) + u
\left[ \frac{1}{\beta} \sum_{\nu^{\prime}_n} D( i \nu^{\prime}_n )
+
q_{EA} \right] - 2 y q_{EA}^2 D(-i \nu_n) \nonumber \\
&&~~- \frac{2 y q_{EA}}{\beta} \sum_{\nu^{\prime}_n} D(i
\nu^{\prime}_n) D(- i\nu_n - i\nu^{\prime}_n) - \frac{2 y}{3
\beta^2} \sum_{\nu^{\prime}_n, \nu^{\prime\prime}_n}
D(i\nu^{\prime}_n) D(i\nu^{\prime\prime}_n) D(- i\nu_n -
i\nu^{\prime}_n-i\nu^{\prime\prime}_n)=0. \label{rsy10}
\end{eqnarray}
Upon consideration of the saddle point equation for $D(0)$ one
initially finds a number of additional term associated with the
coupling of $D(0)$ to the $q_{ab}$. However, our parameterization
in (\ref{rsy4}) was chosen judiciously, and has the feature that
all these additional terms vanish upon using (\ref{rsy8}); so,
the result (\ref{rsy10}) applies {\em also} for $\nu_n = 0$.

Let us also note the complete expression for the free energy
density, obtained by inserting (\ref{rsy4}) and (\ref{rsy10})
into (\ref{rsy2}):
\begin{eqnarray}
{\mathcal F}/n &=& \frac{q_{EA} r}{\kappa} + \frac{y^2 q_{EA}^5}{5
\kappa} + \frac{1}{\beta \kappa} \sum_{\nu_n} ( \nu_n^2 + r) D(i
\nu_n) - \frac{\kappa}{3 \beta} \sum_{nu_n} D^3 (i \nu_n) +
\frac{u}{2} \left[ \frac{1}{\beta} \sum_{\nu_n} D( i\nu_n ) +
q_{EA} \right]^2 \nonumber \\
&&~~-  \frac{y q_{EA}^2}{\beta} \sum_{\nu_n} D(i\nu_n) D(-i\nu_n)
- \frac{2 y q_{EA}}{3 \beta^2} \sum_{\nu_n, \nu^{\prime}_n}
D(i\nu_n) D(i\nu^{\prime}_n) D(- i\nu_n - i\nu^{\prime}_n)
\nonumber
\\ &&~~~~~ - \frac{ y}{6 \beta^3} \sum_{\nu_n, \nu^{\prime}_n,
\nu^{\prime\prime}_n} D(i\nu_n) D(i\nu^{\prime}_n)
D(i\nu^{\prime\prime}_n) D(- i\nu_n -
i\nu^{\prime}_n-i\nu^{\prime\prime}_n). \label{rsy11}
\end{eqnarray}

We are now left with the task of solving the saddle-point
equations (\ref{rsy8}) and (\ref{rsy10}) for $q_{EA}$ and
$D(i\nu_n)$, and inserting the result in (\ref{rsy11}). This is
clearly a daunting task, and we will be satisfied in describing
the $T \rightarrow 0$ limit to first order in $y$. This is
similar in spirit to the large $S$ expansion of
Sections~\ref{sec:largeS}, \ref{Thermo}, and we expect that
higher-order corrections in $y$ will not modify the nature of the
low $T$ limit.

First, we consider the case $y=0$. Here a complete analytical
solution is possible, and was presented in Ref.~\onlinecite{rsy}.
We have at $y=0$:
\begin{eqnarray}
q_{EA}^{0} &=& \frac{1}{\beta \kappa} \sum_{\nu_n} |\nu_n| -
\frac{r}{\kappa u} \nonumber \\
D^0 (i\nu_n) &=& - \frac{|\nu_n|}{\kappa} \nonumber \\
{\mathcal F}^0 (T)/n &=& \frac{2}{3 \beta \kappa^2} \sum_{\nu_n}
|\nu_n|^3 - \frac{r^2}{2\kappa^2 u}\nonumber \\
&=& {\mathcal F}^0 (0)/n - \frac{4 \pi^3 T^4}{45 \kappa^2}.
\label{rsy12}
\end{eqnarray}
We observe that the free energy density behaves as $T^4$, while
the specific heat $\sim T^3$ as $T \rightarrow 0$. Notice also
that positivity of $q_{EA}^{0}$ requires an upper bound on $r$,
which we have assumed to hold.

Before considering explicit corrections in powers of $y$, we make
an observation that is valid to all orders in $y$. The solution
for $D(i\nu_n)$ in (\ref{rsy12}), when analytically continued to
real frequencies, $\omega$, has an imaginary part which vanishes
linearly in $\omega$ at small $\omega$. We now show that this
conclusion holds to all orders in $y$; the constraint
(\ref{rsy8}) will play a key role in establishing this result.
Let us write $D(\omega)  = D(0) + i D_1 \omega + \ldots$ for small
$\omega$, where $D(0)$ and $D_1$ are some real constants.
Inserting this in (\ref{rsy10}) and evaluating it at $T=0$ for
small $\omega$, we note that the last two terms in (\ref{rsy10})
have imaginary parts which vanish as $\omega^3$ and $\omega^5$.
Keeping only the leading $\omega$ dependence of the imaginary
part, we obtain the simple expression
\begin{equation}
-2 i \kappa D(0) D_1 \omega - 2 y q_{EA}^2 i D_1 \omega = 0.
\label{rsy12a}
\end{equation}
From (\ref{rsy8}) we see that this condition is satisfied, and so
$D_1$ can be non-zero.

Now, we consider explicit first-order corrections in $y$: we will
see that this leads to terms in the thermodynamics which vanish
more slowly as $T \rightarrow 0$. We can easily use (\ref{rsy8})
and (\ref{rsy10}) to determine the corrections to $D(i\nu_n)$ and
$q_{EA}$ to linear order in $y$; however, these are not needed
here as the shift in the free energy due to such corrections will
only appear at order $y^2$, because the free energy is at a saddle
point. Indeed, to obtain the free energy correct to first order in
$y$, we need only insert (\ref{rsy12}) into (\ref{rsy11}). It is
then quite easy to see that the free energy will have a term of
order $y T^2$, and that the coefficient of this term will be
non-universal and dependent upon the nature of the high energy
cutoff. The required term comes from the $T$ dependence of
$q_{EA}^0$, which from (\ref{rsy12}) is seen to be
\begin{equation}
q_{EA}^0 (T) = q_{EA}^0 (0) - \frac{\pi T^2}{3
\kappa}.\label{rsy13}
\end{equation}
Upon inserting (\ref{rsy13}) and (\ref{rsy12}) into (\ref{rsy11})
we will now obtain numerous terms in which the above $T^2$ term
multiplies $T$-independent, cutoff-dependent terms coming from
the upper bounds in the summations over the $D (i\nu_n)$. As the
relative values of these contributions will depend upon the
nature of the cutoff, there is no general reason for them to
cancel against each other. Hence we obtain a $T^2$ contribution
to ${\cal F}$ and a linear $T$ term in the low $T$ specific heat.

\endgroup


\end{document}